\documentclass[]{aastex62}

\usepackage{longtable}
\usepackage{graphicx}
\usepackage{ulem}
\usepackage{verbatim}

\newcommand{\kms}{km s$^{-1}$}
\newcommand{\solarmass}{M$_\odot$}

\newcommand{\jyperb}{Jy Beam$^{-1}$}
\newcommand{\mtcn}{CH$_3$CN}
\newcommand{\mtttcn}{CH$_3^{13}$CN}
\newcommand{\mtnl}{CH$_3$OH}

\submitjournal{ApJ}
\shorttitle{Active Accretion around S255IR~SMA1}
\shortauthors{Liu, Su, Zinchenko, et al.}

\begin{document}

\title{ALMA View of the Infalling Envelope around a Massive Protostar in S255IR~SMA1}
%\title{ALMA Observations of the Active {\bf Infall in the Massive Star Forming Core S255IR~SMA1}}

\correspondingauthor{Sheng-Yuan Liu}
\email{syliu@asiaa.sinica.edu.tw}

\author{Sheng-Yuan Liu}
\affil{Institute of Astronomy and Astrophysics, Academia Sinica, 11F of ASMAB, AS/NTU No.1, Sec. 4, Roosevelt Rd, Taipei 10617, Taiwan}

\author{Yu-Nung Su}
\affil{Institute of Astronomy and Astrophysics, Academia Sinica, 11F of ASMAB, AS/NTU No.1, Sec. 4, Roosevelt Rd, Taipei 10617, Taiwan}

\author{Igor Zinchenko}
\affil{Institute of Applied Physics of the Russian Academy of Sciences, 46 Uljanov str., 603950, Nizhny Novgorod, Russia}

\author{Kuo-Song Wang}
\affil{Institute of Astronomy and Astrophysics, Academia Sinica, 11F of ASMAB, AS/NTU No.1, Sec. 4, Roosevelt Rd, Taipei 10617, Taiwan}

\author{Dominique M.-A. Meyer}
\affil{Institut f\" ur Physik und Astronomie, Universit\" at Potsdam, Karl-Liebknecht-Strasse 24/25, 14476 Potsdam, Germany}

\author{Yuan Wang}
\affil{Max Planck Institute for Astronomy, K\"onigstuhl 17, 69117, Heidelberg, Germany}

\author{I-Ta Hsieh}
\affil{Institute of Astronomy and Astrophysics, Academia Sinica, 11F of ASMAB, AS/NTU No.1, Sec. 4, Roosevelt Rd, Taipei 10617, Taiwan}

\begin{abstract}

The massive {young stellar object S255IR~NIRS3 embedded in the star forming core SMA1} has been recently observed with a luminosity burst, which is conjectured as a disc-mediated variable accretion event. In this context, it is imperative to characterize the gas properties around the massive young stellar object. With this in mind, we carried out high angular resolution observations with the Atacama Large Millimeter and submillimeter Array and imaged the 900 $\mu m$ dust continuum and the \mtcn\ $J$=19$-$18 $K$=0$-$10 transitions of S255IR~SMA1. The integrated \mtcn\ emission exhibits an elongated feature with an extent of 1800 au in the northwest-southeast direction at a position angle of 165 degree, which is nearly perpendicular to the bipolar outflow. We confirm the presence of dense (a few $\times 10^{9}$ cm$^{-3}$) and hot ($\sim$ 400~K) gas immediately surrounding {the central protostar}. The \mtcn\ emission features a velocity gradient along the elongated ridge and by modelling the gas kinematics based on features in the position-velocity diagram, we infer that the gas is best described by a flattened rotating infalling envelope (or pseudo-disc). A mass {infall} rate of a few $\times$ 10$^{-4}$ \solarmass\ per year is derived. If there exists a putative Keplerian disc directly involved in the mass accretion onto the star and jet/outflow launching, it is likely smaller than 125~au and unresolved by our observations. We show qualitative resemblances between the gas properties (such as density and kinematics) {in 255IR~SMA1} inferred from our observations and those in a numerical simulation particularly tailored for studying the burst mode of massive star formation.

\end{abstract}


\section{Introduction} \label{sec:intro}

Characterizing the mass accretion phenomenon immediately surrounding the central young star remains particularly as an active research area for our understanding of the formation processes of massive stars \citep{Zinnecker07}, those stars with masses greater than 8 $M_\odot$.
From the theoretical perspective, circumstellar discs around young stellar objects (YSOs) play a vital role in gauging the accretion as well as launching the jets and molecular outflows, through which angular momentum can be transported out and removed.
Such disc-outflow interaction has been commonly observed in low mass protostellar systems.
For young massive stars, the presence of disc-outflow serves another critical function.  
By producing holes in the surrounding envelope through outflows, intense radiation from the central newborn massive star can escape, hence reducing the intense radiation pressure, conventionally thought of as a barrier in massive star formation, which may otherwise prevent the gas accretion \citep[e.g.][]{Krumholz09, Klassen16}.
It is therefore anticipated that like their low-mass counterparts, massive YSOs accrete through similar disc-outflow interaction as a scaled-up version of the standard disc-envelope paradigm \citep[e.g.][]{Cesaroni07, Zinnecker07, Beltran16}. 

Observationally, there have been, however, challenges in identifying and studying the putative disc structures around massive YSOs.
Such objects are generally rare and thus at far distances.
They are furthermore deeply embedded in their parental clouds.
Careful choices of observing wavelengths and tracers are imperative for peeling through the encompassing cloud and high angular resolutions are required for resolving the intricate features.
Thanks to the supreme sensitivity and spatial resolution provided by modern facilities, particularly the Atacama Large Millimeter and submillimeter Array (ALMA), there appears mounting, direct and indirect, observational evidence suggesting that flattened rotating structures, sometimes (Keplerian-like) discs, prevail around newborn massive stars.

For early O-type stars (M $>$ 20 \solarmass\ or L $>$ 10$^5$ L$_\odot$), rotating toroids of 10$^4$ au in size were often observed \citep{Cesaroni07}. 
Toward the massive YSO G31.41+0.31, for example, ALMA observations at a resolution of 0\farcs22 (or $\sim$ 1700 au) revealed (linear) velocity gradients in \mtcn\ and CH$_3$OCHO across the core with progressively larger gradients in higher energy lines, suggesting gas rotation speeding up toward the center \citep{Beltran18}.
Meanwhile, the widespread inverse P-Cygni profiles in all observed lines toward its continuum peak indicate the presence of infall, which show signs of acceleration also toward the center based on the observed velocity in transitions with different energies.
A high infall rate at the level of $3 \times 10^{-2}$ \solarmass ($\Omega/4\pi$) was estimated \citep{Beltran18}.
Emerging evidence further reveals the presence of outflow and Keplerian-like discs around some O-type stars. Examples include AFGL~4176 \citep{Johnston15}, G11.92-0.61MM1 \citep{Ilee16}, G17.64+0.16 \citep{Maud18, Maud19}, and IRAS 16547-4247 \citep{Zapata19}.
The disc extents are often found to be on the thousand-au scale, although this might be an observational bias due to the angular resolutions.

For late O- to early B-type stars with masses of 8--20 M$_{\odot}$, (Keplerian-like) disc structures 
have also been found in recent year to coexist and associate with bipolar molecular outflows.
The massive YSO system IRAS~20126+4104 perhaps had received the most attention.
\cite{Keto10} suggested that models including the presence of a warm, dense, and rapidly rotating disc reproduces the observed molecular line spectra, continuum images, and spectral energy distribution more accurately than models without any disc \citep{Keto10}.
The disc was further shown to be stable against gravitational fragmentation through observations of the \mtcn\ emission \citep{Chen16}.
Most recently, efforts are made in detecting or differentiating the (inner) rotating-supported (Keplerian) disc and the (outer) rotating and infalling envelope as those have been seen in low-mass YSO systems.
Toward the YSO system G328.2551-0.5321, which has a bolometric luminosity of $1.3 \times 10^4$ L$_{\odot}$, for example, \citet{Csengeri18} found CH$_3$OH emission tracing the centrifugal barrier where gas from the collapsing envelope likely gets shocked and transition into an accretion disc.
\citet{Zhang19}, on the other hand, found different molecular emission tracing distinctly the inner disc and the outer envelope in G339.88-1.26, another massive YSO similar to the former.

Despite all the above efforts, it is fair to state that more observations and characterization of such discs/toroids remain strongly desired. 
Particularly important is high angular resolution observations for further resolving spatially the structures surrounding massive YSOs.
The S255IR region, at a distance of 1.78~kpc \citep{Burns16}, harbors a near-IR cluster as well as a cluster of UC H{\scriptsize II} regions \citep[e.g.]{Ojha11}. 
{\citet{Tamura91} resolved the the cluster into 32 near infrared sources. 
Meanwhile,  they found a prominent (northeast-southwest oriented) bipolar infrared reflection nebula (IRN) illuminated by the source NIRS3.}
The massive gas reservoir with a substantial total bolometric luminosity of a few 10$^{4}$ L$_\odot$ pointed S255~IR to a region with active embedded formation of massive stars.
Higher angular resolution imagining with ALMA revealed a couple pairs of outflows associated with two dust continuum cores SMA1 and SMA2 previously identified with the Submillimeter Array (SMA). One pair of outflow with a wide opening angle, {in the same direction of the above-mentioned IRN,} is associated with SMA1 while another well collimated outflow pair centered at SMA2 \citep{Zinchenko18}.
Perpendicular to the bipolar molecular outflow at SMA1, a rotating, possibly disc-like, structure has been reported \citep{Wang11, Zinchenko15}.  
\citet{Wang11} detected CH$_3$OCHO emission with a velocity gradient perpendicular to the outflow.
\citet{Zinchenko15} further showed a similar velocity gradient seen in \mtcn\ and \mtnl\ emission around SMA1.
Recently, a flaring event of the 6.7~GHz class II methanol maser in the S255IR region was recorded by \citet{Fujisawa15} and followed up in high angular resolution by \citet{Moscadelli17}.
{Subsequent observations revealed the burst event being associated with the young stellar object NIRS3 in the near infrared and optical bands \citep{Caratti17}. Temporal variability was also detected in the radio \citep{Cesaroni18} and toward the coinciding star forming core SMA1 at the submillimeter band \citep{Liu18}.}
The observed brightening in the continuum, maser, as well as atomic/molecular lines was attributed to very likely a disc-mediated active accretion event \citep{Caratti17, Liu18}.

The tangible evidence of the rotational motion around the luminosity burst object makes {S255IR~NIRS3/SMA1} a prime opportunity for investigating the circumstellar structure around a massive YSO in its active (accretion) phase.
% elusive circumstellar disc
We present in this paper results from our ALMA Band~7 sub-arcsecond observations of the \mtcn\ transitions toward SMA1. 
We describe in Section~2 the observational setup and data reduction procedures, and present in Section~3 the results, including the gas mass, column density, temperature, and kinematics inferred from the analysis of the \mtcn\ transitions, and the radiative transfer modeling experiments. 
We discuss in Section~4 the implications of our findings, and finally outline our conclusion in Section~5.

\section{Observations} \label{sec:obs}

We carried out our observations with the ALMA toward S255IR SMA1 under the project \#2015.1.00500.S in the Cycle~4 science operation. 
These ALMA Band~7 observations include three epochs, with the first two (2016 April 21 and 2016 September 09) being introduced in \citet{Zinchenko17} and the third one (2017 July 20) in \citet{Liu18}.
There were overall at least 39 antennas online with projected baselines ranging between 12~m and $\sim$ 3~km.
Four frequency-division-mode (FDM) spectral windows introduced in \citet{Zinchenko17} were employed by the observations. 
The window centered around 349.0~GHz with a bandwidth of 937.5 MHz and 3840 spectral channels yielded a spectral resolution of 0.488~MHz (0.42 km s$^{-1}$) after the online Hanning-smooth. 
This frequency coverage include not only the newly discovered methanol (CH$_3$OH) maser feature reported by \citet{Zinchenko17} but also the series of methyl cyanide (CH$_3$CN) $J$=19--18, $K$=$0, 1, ... 10$ rotational transitions with their upper state energy levels ranging from 160~K to nearly 900~K.
The calibrator setups have been detailed in \citet{Zinchenko17} and \citet{Liu18}.

The initial calibration and reduction of the obtained data were carried out with the Common Astronomy Software Applications {(CASA)} \citep{McMullin07} %through manual or pipeline calibration procedures
and delivered by the observatory.
Imaging of the continuum data taken at different epochs were first performed separately. 
We then employed self-calibration for different observational epochs as described in \citet{Zinchenko17} to improve the imaging quality and dynamical range.
The antenna gain solutions were applied to both the continuum data as well as the 349.0~GHz spectral window.
The final 900 \micron\ continuum images and their variability were reported in \citet{Liu18}. 
To facilitate the investigation of the intricate features around the S255IR~SMA1, we concentrate here on the data taken with the extended/high resolution configuration at the 2016 September 09 epoch.
As done in \citet{Liu18}, the resulting images were first made through Briggs weighting with a robust parameter of 0.5 and then tapered to an angular resolution of 0\farcs14.
This angular resolution corresponds to a linear scale of $\sim$ 250 au. The resulting RMS noise in the image cube at our 0.42 \kms\ spectral resolution is 3~m\jyperb. We adopt a systemic velocity of 5.2 \kms\ for the target.

\section{Results} \label{sec:results}

As introduced in Section \ref{sec:intro}, a luminosity burst {at submillimeter band} took place in S255IR~SMA1 in mid 2015. After reaching its climax, there was the evidence of declines in the \mtnl\ masers and submillimeter continuum fluxes \citep{Liu18, Szymczak18}.
Our two epochs of observations, conducted in 2016 September and 2017 July are in the post-burst phase. We present here the data/images obtained in the 2016 September epoch. 

\subsection{Dust continuum, gas column density and mass}

In Figure \ref{fig:mom0} we display in false-color the 900 \micron\ dust continuum emission, which {has a rms noise level of $\sim$ 1~m\jyperb\ and} exhibits its strongest compact emission peaking at R.A. = 06:12:54.013 and Dec. = 17:59:23.05, the nominal position of the S255IR~SMA1 source.
The lower level continuum emission displays an elongation in the general northwest-southeast direction but also with a protrusion pointed towards northeast.
Given the dust and gas are likely closely coupled and the gas temperature is at a level above 300~K toward the peak (see Section 3.3), the peak continuum intensity of 0.237~\jyperb, corresponding`` to a brightness temperature of 121~K, implies that the dust continuum emission, though with a non-negligible opacity, is mostly optically thin at this 0\farcs14 scale.
With the assumption of a dust temperature of 300 K (see Section 3.3),
we translate this peak intensity $F^{\mbox{beam}}_{\nu}$ to {a gas column density $N_{\mbox{\tiny H$_2$}}$ of 9.75 $\times$ 10$^{24}$ cm$^{-2}$} based on Equation A.26 in \citet{Kauffmann08} as shown below

\begin{eqnarray}
N_{\mbox{\tiny H$_2$}} & = & \frac{F^{\mbox{beam}}_{\nu}}{\Omega_{\mbox{\tiny A}} \mu_{\mbox{\tiny H$_2$}} m_{\mbox{\tiny H}} \kappa_{\nu} B_{\nu}(T)}
\end{eqnarray}

where $\Omega_{\mbox{\tiny A}}$ is the beam size of 0\farcs14, $\mu_{\mbox{\tiny{H2}}}$ is the molecular weight per hydrogen molecule of 2.8, $m_{\mbox{\tiny{H}}}$ is the atomic mass of hydrogen, $\kappa_{\nu}$ is the specific absorption coefficient (per mass), and \emph{B}$_{\nu}$(\emph{T}) is the Planck function at a temperature \emph{T} of 300K.
{Here, we adopt $\kappa_{\nu}$ of 0.010 cm$^{2}$g$^{-1}$ by using $\kappa_{\nu} = \kappa_{250 \mu m} ({250 \mu m}/{\lambda})^{\beta}$ with $\kappa_{250 \mu m} = 0.1$ cm$^{2}$g$^{-1}$ \citep{Hildebrand83}, $\beta = 1.8$ \citep{Chen16}, and implicitly a gas-to-dust mass ratio of 100.}
When the source distance of 1.78~kpc is taken into account, it implies enclosed within the central 250~au {a total mass of around 0.36 \solarmass, or equivalently an average gas density of $\sim$ 6.6 $\times$ 10$^{9}$ cm$^{-3}$}.
The integrated fluxes within larger circular apertures with diameters of 0\farcs5 ($\sim$ 900 au), 1\farcs0 ($\sim$ 1800 au), and 1\farcs5 ($\sim$ 2700 au) are 0.613~Jy, 0.932~Jy, and 1.08~Jy, respectively.
With the same assumptions above, the enclose masses within thousand au scale are estimated {to be 0.930 \solarmass, 1.41 \solarmass, and 1.64 \solarmass.} 
These enclosed masses follow roughly $r^{0.5}$, where $r$ is the radius enclosing the volume, hinting a volume density profile scales roughly with $r^{-2.5}$ if a simple spherical geometry is considered.
If a flattened two-dimensional (disc/toroid-like) geometry is assumed, the surface density could instead scale as $r^{-1.5}$ and the corresponding volume density in the inner region likely would be even higher than that quoted above due to the assumed geometry.
We note, though, that the averaged temperature over a larger area would likely be cooler, and possibly certain missing flux may start kicking in, particularly when the angular scale reaches beyond 2\arcsec, the maximum recoverable angular scale of the observations.
Both effects may lead to larger masses at larger volumes and the underlying density profile could hence be somewhat shallower than that inferred above.
{The overall mass estimates, nevertheless, suffer from various uncertainties as discussed in Section 4.4.}

\subsection{Spectral features}

Figure \ref{fig:figwinspec} shows the spectrum spanning over the full spectral window within a synthesized beam centered at the continuum peak.
We identified the most probable carriers of the observed spectral features in both our spectrum and image-cube by cross-referencing Splatalogue \footnote{https://splatalogue.online/}, an online database which consolidates the Jet Propulsion Laboratory Molecular Spectroscopy \citep[JPL\footnote{https://spec.jpl.nasa.gov/},][]{Pickett98} and the Cologne Database for Molecular Spectroscopy \citep[CDMS\footnote{https://cdms.astro.uni-koeln.de/},][]{Muller05} catalogs, and list them (in the order of descending frequency) in Table \ref{tab:trans}.
{The majority of these identified features originate from known interstellar complex organic molecules, including NH$_2$CHO, CH$_3$CHO, CH$_3$OCHO, CH$_3$CH$_2$OH, and (CH$_2$OH)$_2$, with multiple transitions at reasonable excitation energy levels present within the band, although analyzing their excitation is beyond the scope of this work.
The remaining, generally weak, spectral features from our best guess but uncertain carriers are marked with question marks in the Table.}
The \mtnl\ maser newly identified by \citet{Zinchenko17} is clear seen as the strongest feature in the spectrum, with its intensity over 4~\jyperb (or equivalently over 2000~K in brightness temperature).
This maser emission, at the strong peak location at the north-east of the continuum peak, as noted in \cite{Zinchenko17}, reaches $\sim$ 5.9 \jyperb at $\sim$ 0\farcs13 resolution and exceeds equivalently a brightness temperature of 3900 K \citep{Zinchenko17}.

In this work, our primary focus is the \mtcn\ transitions with their quantum numbers running from $J$=19--18, $K$=0 up to $K$=10, which are labeled in Figure \ref{fig:figwinspec}.
Among the detected molecular lines listed in Table \ref{tab:trans}, we highlight with boldface these transitions, whose upper level energies range from 167.7~K to 880.7~K.
To better visualize the line profiles of the \mtcn\ $J$=19--18 $K$ components in detail, we present in Figure \ref{fig:figch3cnspec} the individual $K$=0 to $K$=10 spectra extracted from Figure \ref{fig:figwinspec} but plotted in the velocity space.
The typical {observed} full-width-half-maximum (FWHM) linewidths of the \mtcn\ features are on the order of 8 \kms.
As one can notice from both Figure \ref{fig:figwinspec} and Figure \ref{fig:figch3cnspec}, the \mtcn\ $K$=0 and $K$=1 transitions are blended themselves. 
A few other \mtcn\ lines with higher $K$ are also noticeably blended by interlopers.
Although the identifications of blended lines may not be obvious in this display (Figure \ref{fig:figwinspec}), they become clearer in the position-frequency plot discussed later.
The \mtcn\ $K$=5 and $K$=6 transitions, for example, are blended with the \mtttcn\ transitions, respectively, on the red-shifted and blue-shifted ends.
Despite the line blending, one can spot in most \mtcn\ transitions doubly peaked, asymmetric, (blue-)skewed profiles.
The $K$=8 line is the only exception, where its red-shift end is contaminated by a relatively strong NH$_2$CHO feature.
Doubly peaked blue-skewed profiles, {if not resulted from line-blending,} could be due to two velocity components, or as are often considered, an indication of inflow motion.

With a quick glance over the spectra, one may also find the fact that the intensities of the $K$=3 and even $K$=6 transitions are quite comparable to those of the $K$=0, 1, and 2 components.
Given that the transitions of \mtcn\ with $K=3n$ ($n=1,2,...$) are doubly degenerate as compared to the $K\ne3n$ lines, the $K$=3 and $K$=6 components, for example, in the optically thin regime would presumably appear roughly twice brighter than their neighboring K lines.
The comparable intensities in the observed spectra thus indicate that these lines are optically thick.
A parallel line of evidence for the high optical depth of the lower $K$ lines comes from the observed \mtttcn\ intensities as compared to those of \mtcn.
Although the isotopic ratio between $^{12}$C and $^{13}$C appears to vary with Galactocentric radii with large intrinsic scatters, the value is likely no smaller than 50 \citep{Wilson94, Wilson99} for the S255IR region.
Therefore, the column densities of \mtcn\ and \mtttcn\ and the corresponding peak intensities of their transitions, in the optically thin regime, should display a similar ratio.
The intensity ratios of the observed \mtcn\ $J$=19--18 $K$=0 and $K$=1 lines to those of the $^{13}$C isotopologues, as can be estimated from Fig \ref{fig:figwinspec}, are much smaller than that, hinting again a significant opacity in the \mtcn\ low $K$ lines.
With the continuum baseline taken into account, the low-lying optically thick \mtcn\ transitions appear saturated at an apparent brightness temperature of around 274~K at $\sim$ 0\farcs14 resolution.

\subsection{\mbox{\mtcn}: distribution and temperature}

\begin{figure*}
\plotone{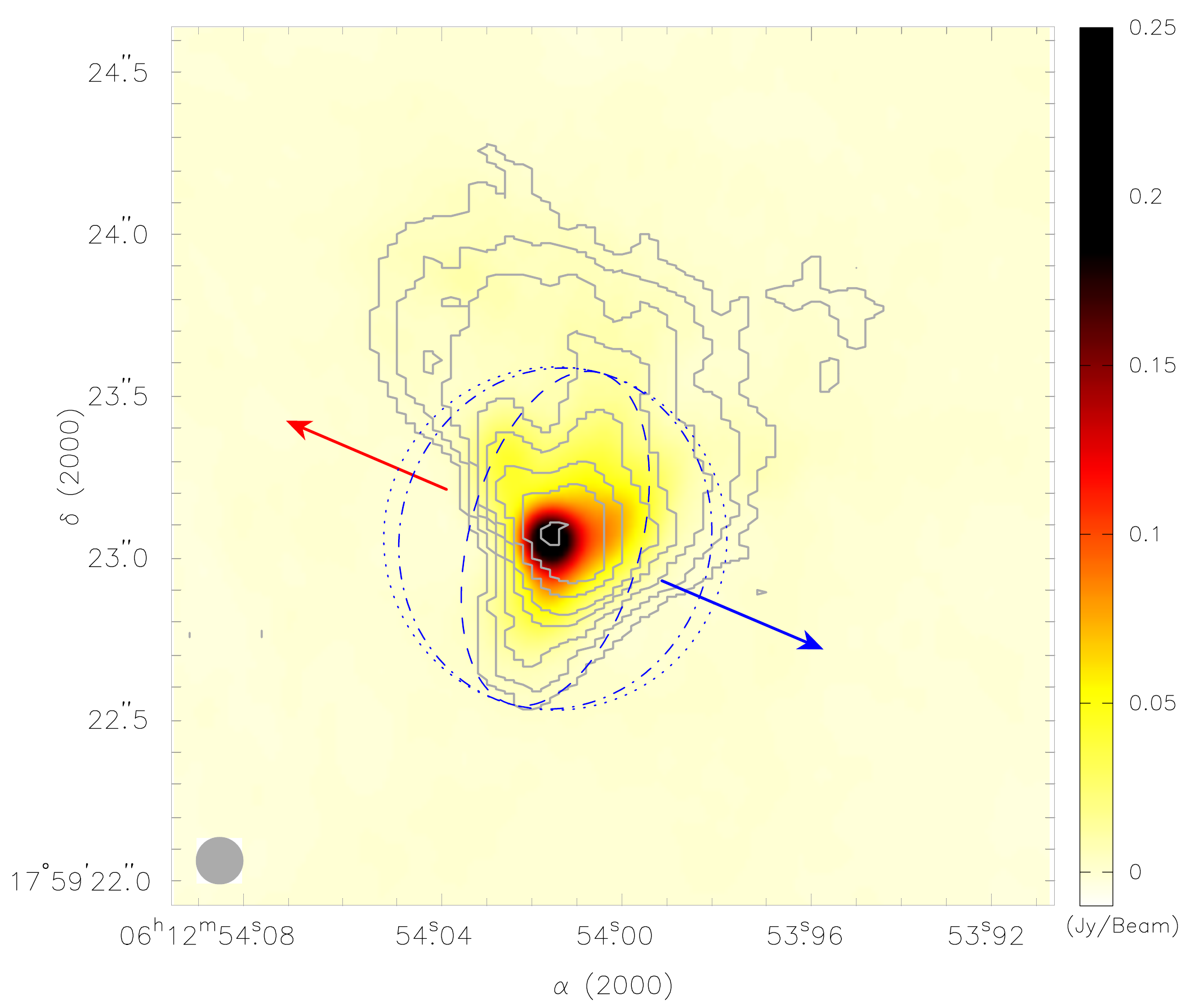}
\caption{\label{fig:mom0} Contours represent the integrated \mtcn\ $J=19-18$ $K$=3 emission overlaid with the 900~$\mu$m continuum in false color scales. The gray circle at the bottom left denotes the synthesized beam for both \mtcn\ and 900~$\mu$m images. Contour levels are 3, 6, 10, 15, 20, 30, 40, and 50 $\times$ 30 mJy Beam$^{-1}$ km s$^{-1}$. The blue/red arrows mark the direction of CO outflows powered by S255IR SMA1 \citep{Zinchenko15}. The dotted, dash-dotted, and dash lines delineate, respectively, the perimeter of an axisymmetric circumstellar structure with a diameter of 1\arcsec, a P.A. of 165 degree, and an inclination angle of 0 degree, 25 degree, and 60 degree.}
\end{figure*}

\begin{figure*}
\epsscale{1.12}
\plotone{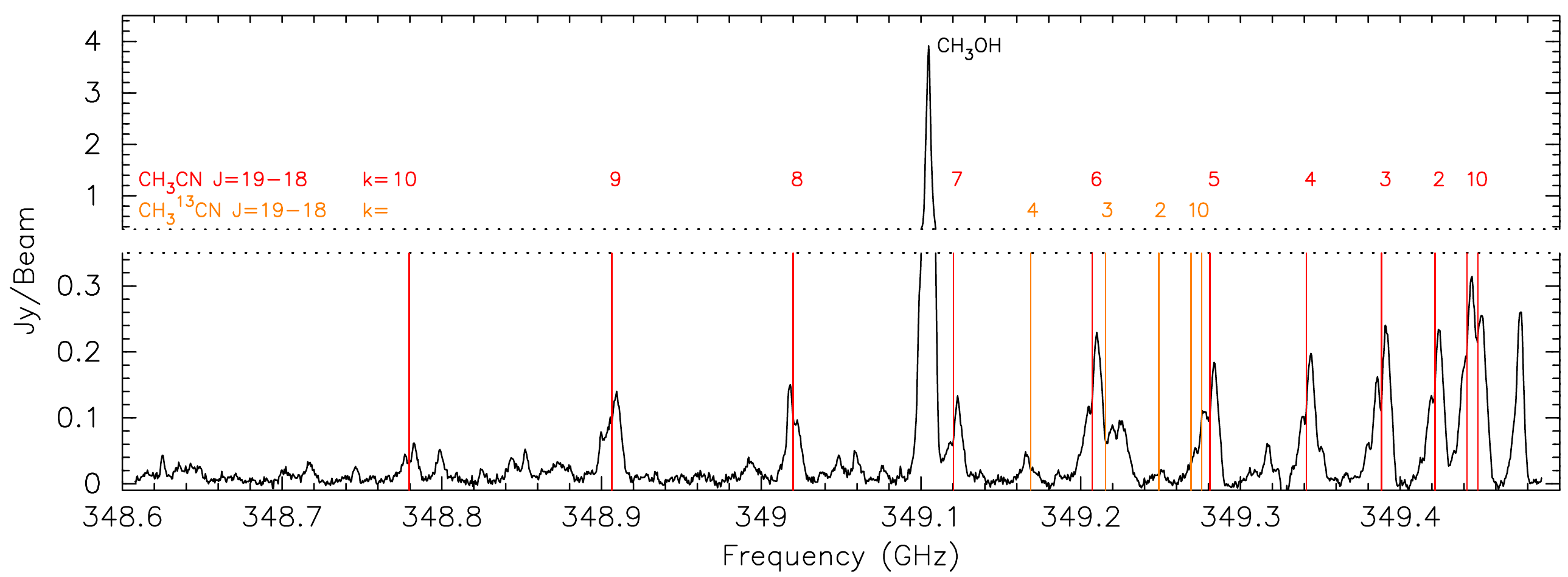}
\caption{\label{fig:figwinspec} The full spectrum toward the S255IR SMA1 continuum peak at 0\farcs14 resolution. The \mtcn\ $J$ = 19$-$18, $K$=0 to $K$=10 lines and the \mtttcn\ $J$ = 19$-$18, $K$=0 to $K$=4 liens are marked by the red and orange vertical lines.}
\end{figure*}

\begin{figure*}
\epsscale{1.0}
\plotone{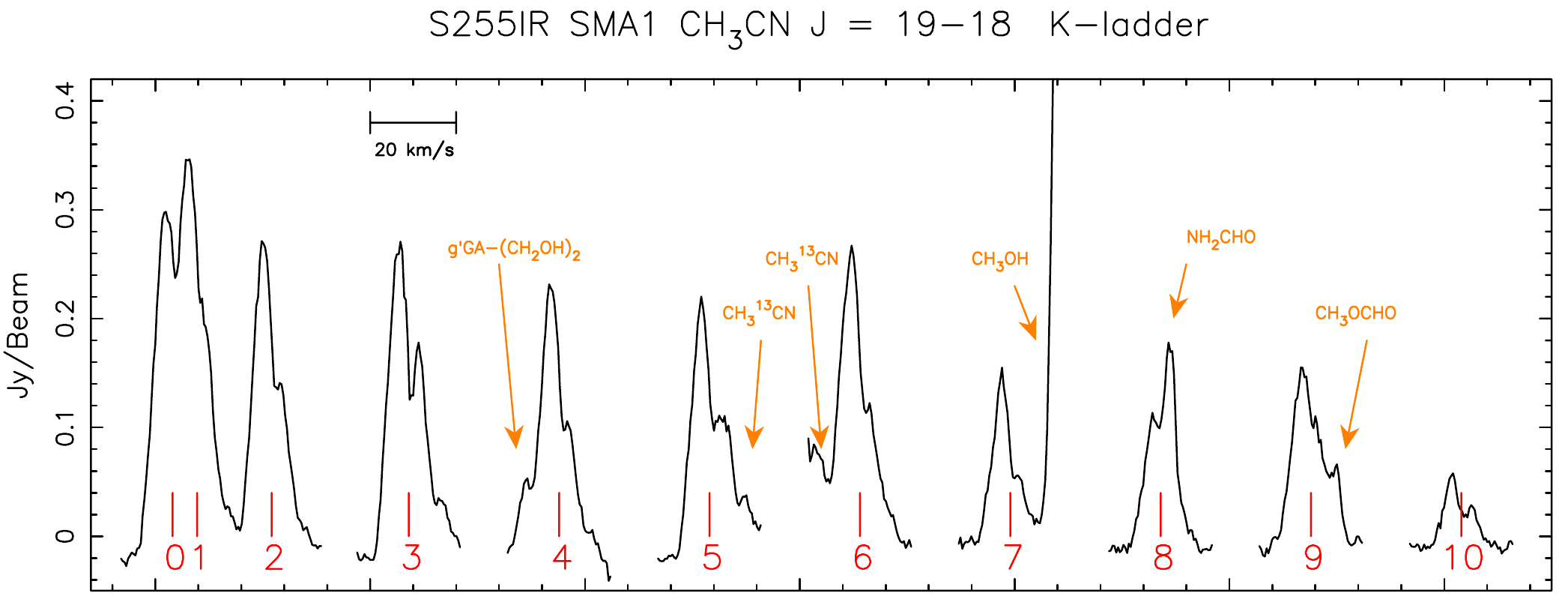}
\caption{\label{fig:figch3cnspec} The spectra of the \mtcn\ $J$ = 19$-$18, $K$=0 to $K$=10 lines toward the S255IR SMA1 continuum peak again at 0\farcs14 resolution. The horizontal axis is in velocity scale as indicated by the horizontal bar above the $K$=3 transition.}
\end{figure*}

The general distribution of the \mtcn\ emission is depicted in the integrated intensity map of the \mtcn\ $J$=19--18 $K$=3 transition shown by the contours in Figure \ref{fig:mom0}.
This integrated emission, as shown in the figure, is resolved with its strongest peak closely coinciding with the continuum peak. 
The map exhibits a prominent elongation of \mtcn\ emission, which has an extent of around 1\arcsec\ (or 1800~au) and a position angle (P.A., measured from north towards east) of 165 degree.
This P.A. is nearly perpendicular to the axis of the large scale bipolar molecular outflows seen by the SMA and the IRAM~30M telescope with a P.A. = 67 degree \citep{Zinchenko15} as marked by the blue and red arrows in Figure \ref{fig:mom0}, and is closely orthogonal to the small scale H$_2$O maser jet structure (with a P.A. = 49 degree) imaged through {the Very Long Baseline Interferometric (VLBI) observations by \citet{Goddi07} and \citet{Burns16} using the Very Long Baseline Array (VLBA) and the VLBI Exploration of Radio Astrometry (VERA), respectively.}

\begin{figure*}
\epsscale{1.0}
\plotone{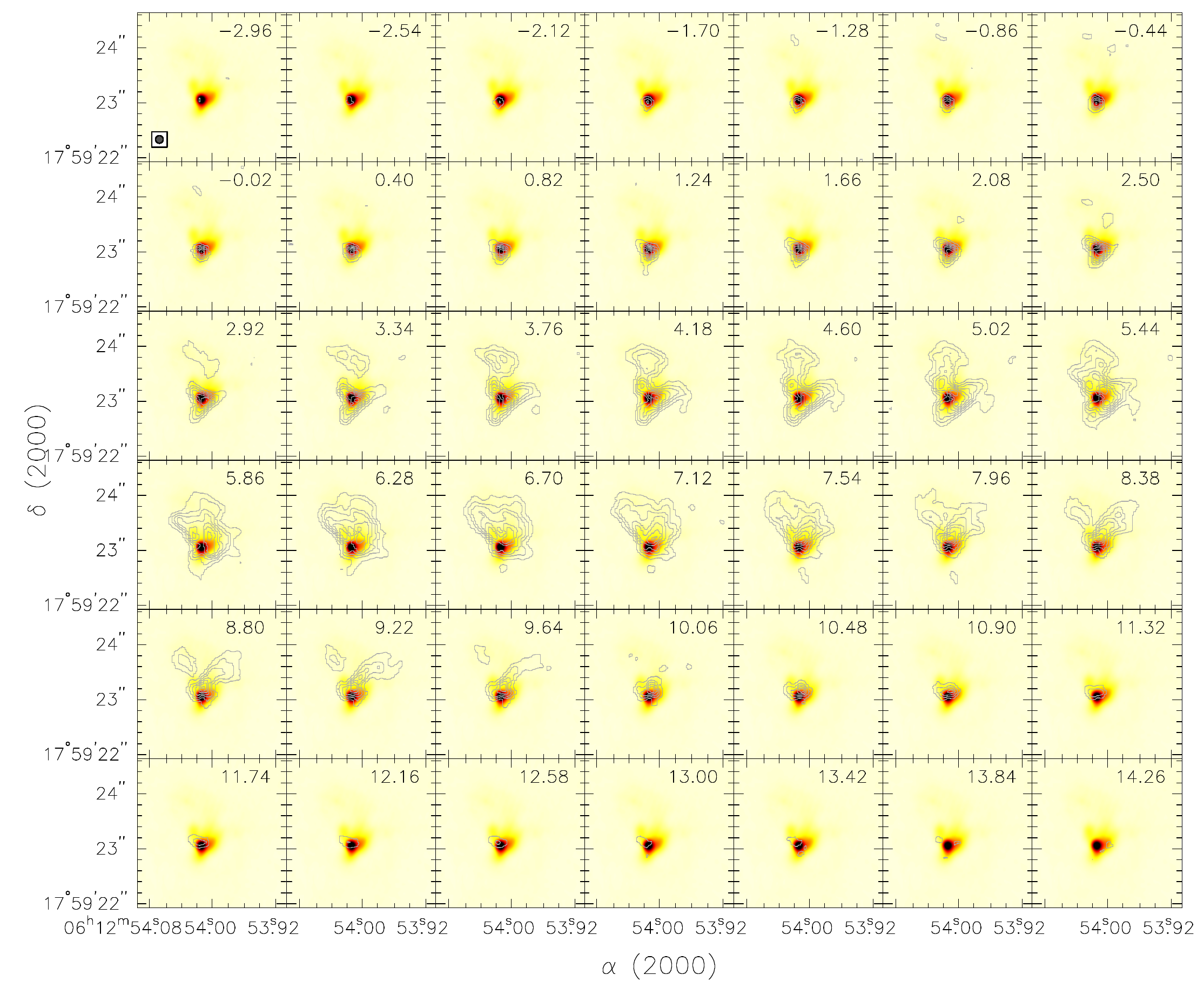}
\caption{\label{fig:channel} Channel maps of the \mtcn\ $J$=19$-$18 $K$=3 component toward S255IR SMA1. The local standard of rest (LSR) velocity (in km s$^{-1}$) of each channel is indicated in the upper right corner of each panel. The LSR velocity of the system is 5.2 km s$^{-1}$. Contours are at 3, 8, 13, 18, 23, 28, and 33 $\times$ 6 mJy Beam$^{-1}$ {($= 2 \sigma$)}. For each panel, the overlaid color scales are the 900~$\mu$m continuum emission. The synthesized beam is 0$\farcs$14 for both the 900~$\mu$m continuum and the \mtcn\ images and is shown at the bottom left corner of the first panel.}
\end{figure*}

The full \mtcn\ emission morphology, however, is more complicated.
Similar to the continuum emission, there exists weaker level of \mtcn\ emission extending toward the northeast direction. 
The channel maps of the \mtcn\ $J$= 19--18, $K$=3 transition, shown in Figure \ref{fig:channel}, further detail the distribution of the \mtcn\ emission around S255IR~SMA1 at different velocities.
From around 3~\kms\ to 9~\kms, extended \mtcn\ appears to be associated with the continuum and also toward the north-east.
Part of the emission on the north-east end at around 3~\kms\ and 5~\kms\ may be co-spatial with the bright masing \mtnl\ emission, which resides in this north-east region.
At the extreme velocities ($< 0$ \kms\ or $> 10$ \kms), the \mtcn\ emission appears very compact in the close vicinity of SMA1.

%\subsubsection{gas temperature}

Since \mtcn\ is a symmetric-top rotor, radiative transitions between different $K$ levels are prohibited, and these levels are therefore populated solely by collisional processes. 
For this reason, the line ratios of these $K$ components are commonly used for probing the gas (kinetic) temperature. 
We perform synthetic spectral profile fitting of the \mtcn\ spectra pixel by pixel following the approach in \cite{Wang10} and \citet{Hung19}.
One additional consideration for the analysis in this study is to take into account the contribution from the non-negligible continuum baseline in the true line brightness temperature.
{In the process we first carry out a simultaneous multiple--Gaussian profile fitting to the set of observed \mtcn\ transitions based on the known frequency separation of these \mtcn\ components, through which the systemic velocity and linewidth of the \mtcn\ emission get determined. 
The process further incorporates a set of parameters, including the \mtcn\ column density, gas (kinetic) temperature (which is assumed to be the same as the dust temperature), beam filling factor, and (dust) continuum optical depth for generating the synthetic spectra. 
The local thermal dynamical equilibrium (LTE) condition is assumed for setting the populations of different $K$ levels.
By varying the input parameters and minimizing the difference ($\chi^2$) between the observed and the synthetic spectral profiles, we found the optimal set of parameters.
The parameter space we considered is the following: 10$^{14}$~cm$^{-2}$ -- 10$^{19}$~cm$^{-2}$ for the \mtcn\ column density, 3~\kms\ -- 6~\kms\ for the LSR velocity, 3~\kms\ -- 6~\kms\ for the linewidth, 10~K -- 500~K for the gas temperature, 0.02 -- 1.0 for the filling factor,  and 0.0 -- 1.0 for the (dust) continuum optical depth.
We limited our fitting exercise to the pixels where the \mtcn\ $K$=3 transition is detected above 3$\sigma$ noise level.}

\begin{figure*}
\epsscale{1.1}
\plotone{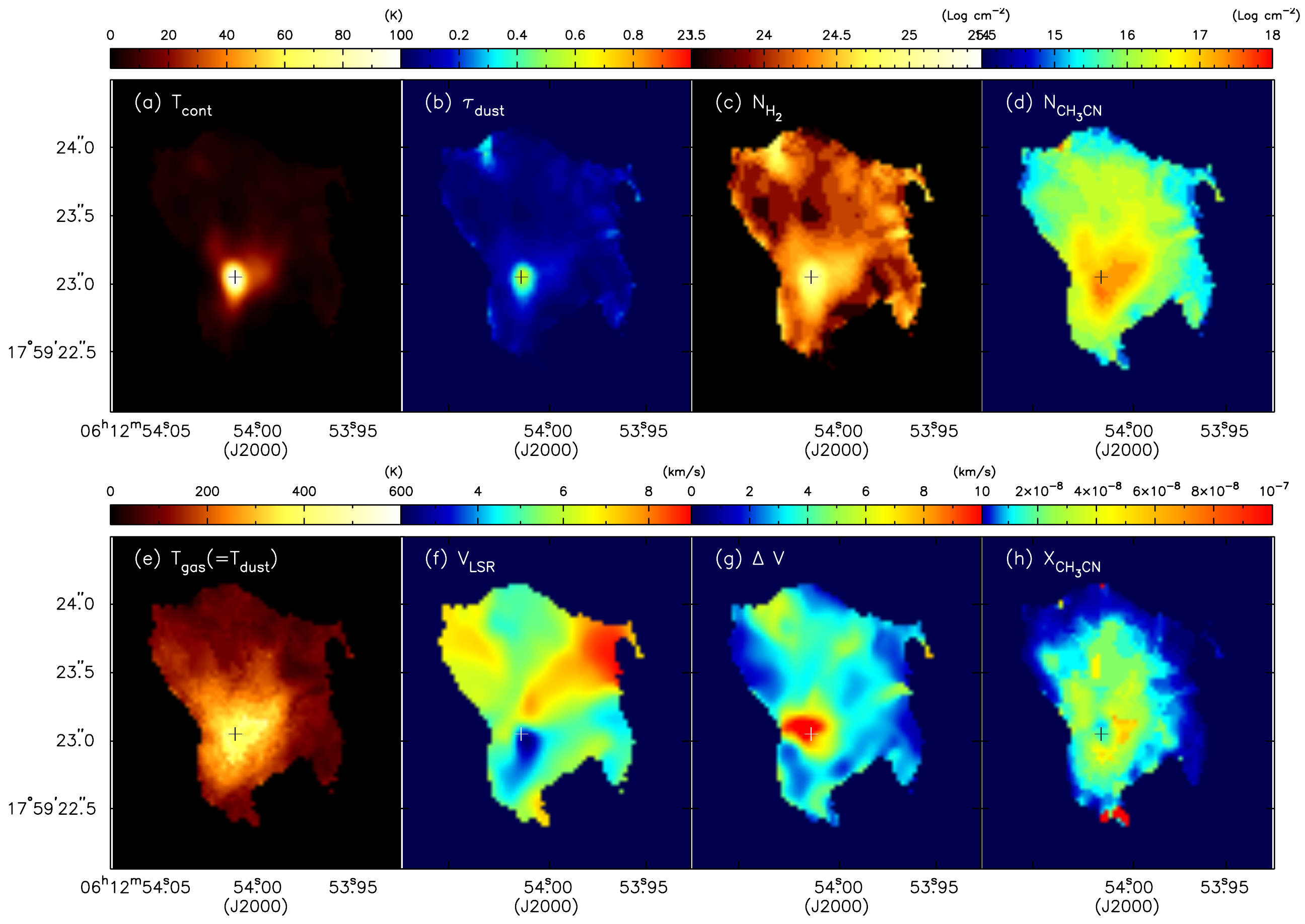}
\caption{\label{fig:fitplot} Images of the spectral profile fitting parameters, which include in panel (a) the continuum brightness temperature, (b) the dust continuum opacity, (c) the molecular gas column density, (d) the \mtcn\ column density, (e) the \mtcn\ temperature, (f) the \mtcn\ velocity field, (g) the \mtcn\ velocity dispersion (linewidth) and (h) the \mtcn\ fractional abundance. The white/black cross denotes the position of {S255IR~NIRS3/SMA1}.}
\end{figure*}

Figure \ref{fig:fitplot} presents the parameters inferred from the synthetic spectral profile fitting.
Panel (a) shows the dust continuum brightness temperature, which is in essence equivalent to the continuum emission map shown in Fig.~1.
Panel (d) displays the distribution of the \mtcn\ column density.
The dominant feature is a high column density ridge elongated in the northwest-southeast direction surrounding SMA1.
From panel (e), we find that the (\mtcn) gas temperature peaks toward SMA1 at more than 400~K and remains hot in the high column density regions.
While the temperature gradually falls off toward the peripheral areas, within most of the fitted regions with an extent of 1\farcs\ or 1800~au the temperature is greater than 100~K.

Based on the continuum emission brightness and the inferred gas/dust kinetic temperature, the dust optical depth , $\tau_{\mbox{\scriptsize dust}}$, is constrained and shown in panel (b). 
A clump with elevated opacity is apparent and coinciding with SMA1, with the dust continuum emission being mostly optically thin throughout the region when averaged over the synthesized beam.
The inferred dust optical depth, based on equation A.9 and A.10 in \citet{Kauffmann08}, can be translated into the molecular hydrogen column density.
We present the column density map in panel (c), which is fundamentally equivalent to panel (b) 
but scaled into the (log) column density unit.
Toward the SMA1 peak at 250~au scale, we reach a molecular gas column density of $1.4 \times 10^{25}$ cm$^{-2}$, consistent with the column density derived in Section 3.1 within 50\%.
As noted in Section 3.1, the derived mean gas density is {on the order of 10$^{9}$ -- 10$^{10}$ cm$^{-3}$}.
For the \mtcn\ $J$=19--18 transitions, their critical densities for thermalization are around 10$^{7}$ cm$^{-3}$.
The ``effective excitation densities'' could be further reduced in the case of optically thick emission \citep{Shirley15}.
The LTE condition used in the spectral profile fitting is therefore a valid {self-consistent} assumption.

\begin{figure*}
\epsscale{1.0}
\plotone{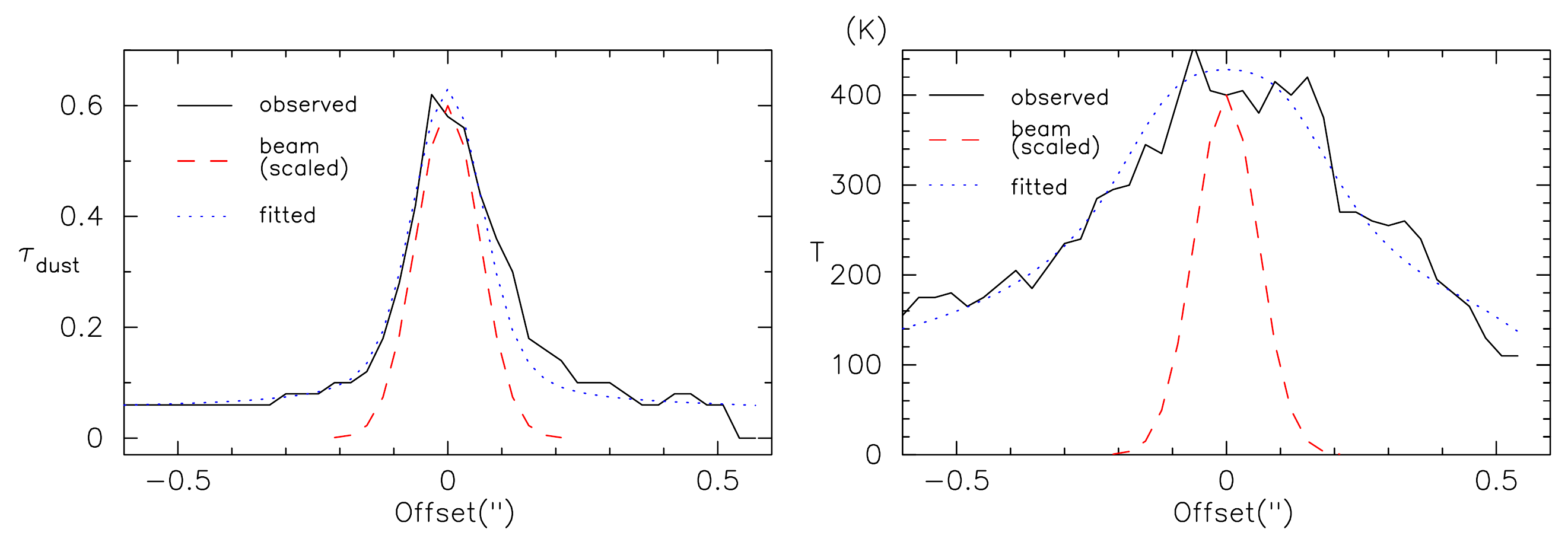}
\caption{\label{fig:profplot} The profiles of physical parameters, including in the left panel the dust optical depth (which is proportional to the gas column density) and in the right panel the gas temperature of the S255IR SMA1 region.}
\end{figure*}

We attempt to establish empirically the radial profiles of the molecular gas column density and the (\mtcn) gas temperature by extracting their values along the major axis from Figure \ref{fig:fitplot} panels (b) and (e) and plotting them in Figure \ref{fig:profplot}.
Both profiles appear resolved as compared to the beam profile also shown in the same figure.
We assume the gas column density, or equivalently the gas surface density, varies as $r^{-p}$ where $r$ is the radius.
We find that the surface density profile, when convolved with the beam profile, is best fitted with $p=1.1$, as shown in the left panel of Figure \ref{fig:profplot}.
For the gas temperature, we adopt the \mtcn\ temperature, which appears plateaued near the central YSO but vary with $r^{-0.7}$ beyond around 200~au.
The (beam-convolved) fitted temperature profiles is plotted in Figure \ref{fig:profplot} right panel.

The inferred \mtcn\ column density toward the SMA1 is $3.0 \times 10^{17}$ cm$^{-2}$, thus implying a \mtcn\ fractional abundance of $2 \times 10^{-8}$. 
The same level of fractional abundance is found throughout the region as shown in Figure~\ref{fig:fitplot} panel~(h).
The value is somewhat higher when compared with the \mtcn\ abundance, $10^{-9}$ \citep[and references therein]{Su09}, often found in massive star forming regions.
As \cite{Su09} revealed in the molecular envelope of the H{\scriptsize II} region G5.89, an enhancement of \mtcn\ fractional abundance higher than $10^{-9}$ can occur in the inner and hotter region due to the evaporation of grain icy mantles.
{Indeed, this abundance is perfectly consistent with values found towards a sample of 17 hot molecular cores by \citet{Hernandez14}.}
%Indeed, the slightly higher fractional abundance is within the range inferred for the Sgr B2(N) region \citep{Pols18} and adopted by, for example, \citet{Johnston15} and \citet{Chen16} for gas around massive YSOs.

\subsection{\mbox{\mtcn}: kinematic signatures}

\begin{figure*}
\epsscale{1.1}
\plotone{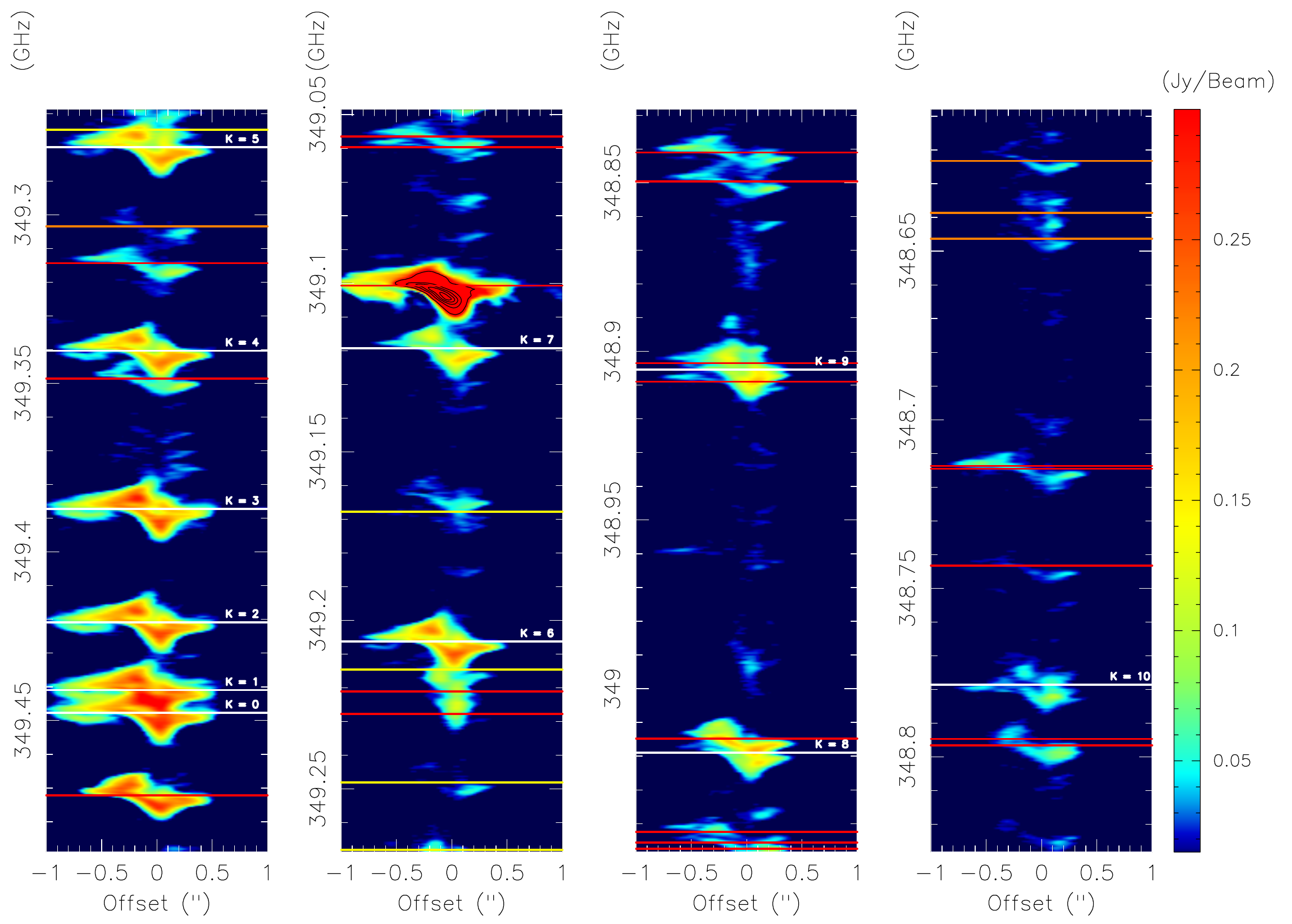}
\caption{\label{fig:figfullpv} The position-frequency diagram for the full spectral window from a cut crossing the S255IR~SMA1 continuum peak at a position angle of 165 deg. The zero offset centers at S255IR~SMA1 with positive offset in the south-east direction and negative offset in the north-west direction. Horizontal lines in white marks the \mtcn\ transitions while those in yellow marks the \mtttcn\ transitions. The rest of the identified and tentatively identified features are highlighted by red and orange horizontal lines.}
\end{figure*}

The \mtcn\ velocity and linewidth (velocity dispersion) signatures are shown in Figure \ref{fig:fitplot} panels (f) and (g), respectively.
A velocity gradient is evident along a stripe from the northwest in red to the southeast in blue.
To illustrate this velocity gradient across SMA1, we present in Figure \ref{fig:figfullpv} the position-frequency (P-F) diagram of spectral features within the entire spectral window made with a cut running through the SMA1 peak at a P.A. of 165 degree.
The spectrum toward the continuum peak shown in Figure \ref{fig:figwinspec} is chiefly the intensity profile along the zero offset pixel in the cut.
It is evident that, as those of \mtcn\ (marked with white lines), all spectral features manifest very similar patterns, which enables better line identification by rejecting random noisy features along just one pixel.
The carrier of these spectral features identified from this figure, including \mtttcn\ and other species, are marked in yellow and red, respectively and also listed in Table \ref{tab:trans}.
{The few features with uncertain carriers noted earlier are further marked in orange.}
Still, there remains some features whose carriers are not identified.

A similar gradient in the general northwest-southeast direction has been reported by \citet{Wang11} and \citet{Zinchenko15}.
\cite{Wang11} detected a velocity gradient in C$^{18}$O and CH$_3$OCHO. 
The extend of the C$^{18}$O emission, spanning nearly 10\arcsec\ scale, was suggested to trace a rotating toroid that got fragmented into a multiple system including at least SMA1 and SMA2.
The CH$_3$OCHO emission, not well resolved by their 1\farcs8 beam, exhibits a linear velocity gradient within 2\arcsec\ scale and was thought to show likely the presence of a rotating and infalling core similar to the toroids described in \citet{Cesaroni07}.
\cite{Zinchenko15} interpreted the velocity gradient, observed at a higher resolution of 0\farcs5 in CH$_3$OH and \mtcn, as a hint of Keplerian rotation within a rotationally-supported disc.
In Figure \ref{fig:fitplot} panel (f), we notice a hint of twisting in the position angle of this velocity gradient from the outside connecting toward the inner region around SMA1.
The region with the highest velocity dispersion, as illustrated by Figure \ref{fig:fitplot} panel (g), is very close to SMA1. 
In addition, a red-shifted component is visible toward the northeast, which is probably related to the red-shifted outflow lobe.

\begin{figure*}
\epsscale{1.1}
\plotone{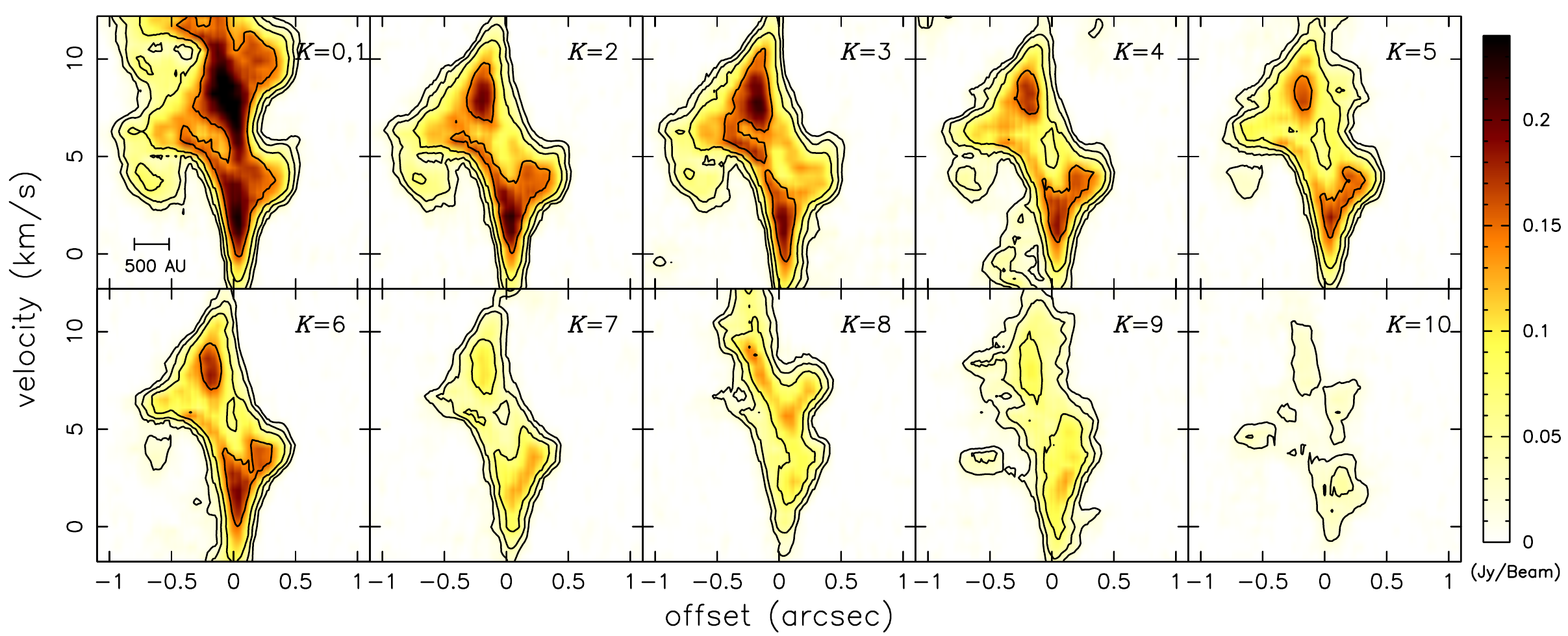}
\caption{\label{fig:figpvplot} The position-velocity diagram of S255IR SMA1 in the \mtcn\ $J = 19-18$ $K$-ladder along the P.A. of 165 deg in both false color and contours.
The contour levels are at 3, 6, 12, and 24 $\times$ 6mJy beam$^{-1}$ {($= 2 \sigma$)}.  The $K$ component is labeled in the upper right corner of each panel and 
the length of a physical scale of 500 au is labeled in the top-left panel. Same as in Fig. \ref{fig:figfullpv}, the zero offset centers at S255IR~SMA1 with positive offset in the south-east direction and negative offset in the north-west direction.}
\end{figure*}

From Figure \ref{fig:figfullpv}, we extract and plot in Figure \ref{fig:figpvplot} the position-velocity (P-V) diagram of the \mtcn\ $J$=19-18, $K$=0 through $K$=10 transitions for further scrutinizing the northwest-southeast velocity gradient across SMA1.
In the first panel, the $K$=0 and $K$=1 components are blended and shown together.
Apparent blending features are also discernible in the $K$=4, $K$=8, and $K$=9 panels.
Two important attributes in the P-V diagram are readily recognized.
First, an overall velocity gradient across the cut is again evident from the blue-shifted gas at the positive (SE) end to the red-shifted gas at the negative (NW) end. 
There is a quite abrupt flip of the velocity near the central position.
Second, while the blended $K$=0 and $K$=1 components appear relatively extended, tracing a smaller velocity gradient, at progressively higher $K$s, the emission extent becomes more compact and the apparent velocity gradient gets steeper. 

What is the origin of this observed velocity gradient pattern?
Since this gradient is closely perpendicular to the general direction of the bipolar outflows, it most plausibly indicates rotation.
Furthermore, as the higher K transitions trace the more compact, hotter, and presumably denser gas closer to the embedded protostellar object, the increasing velocity gradient signifies a spin-up rotation toward the center.
It is tempting to attribute the velocity gradient to the Keplerian rotation of a rotationally supported circumstellar disc around the central YSO.
Plotted in Figure \ref{fig:figpvcmp} panel (a) are, respectively, red, blue, and cyan curves corresponding to Keplerian rotation of (8.04/$sin^2 i$) \solarmass, (3.57/$sin^2 i$) \solarmass, and (0.89/$sin^2 i$) \solarmass, where $i$ represents the inclination angle of the system with $i = 0$ meaning face-on.
For an inclination angle of 25 deg, the red, blue, and cyan line, for example, denote the Keplerian curves for a central mass of 45~\solarmass, 20~\solarmass, and 5~\solarmass, respectively.
For an inclination angle of 60 deg, the red, blue, and cyan line, then stand for the Keplerian curves for a central mass of $\sim$ 10.7~\solarmass, 4.8~\solarmass, and 1.2~\solarmass, respectively.
We plot in Fig \ref{fig:mom0} the perimeters of a axisymmetric circumstellar structure with a diameter of 1\arcsec, a P.A. of 165 degree, and an inclination angle of 0 degree, 25 degree, and 60 degree.
The degree of elongation in our integrated \mtcn\ emission, namely the minor axis of the structure being roughly half in length of the major axis, would suggest an inclination angle of no less than 60 degree.
It is evident that the emission pattern on the red-shifted and blue-shifted sides are not fully symmetric. Meanwhile, the blue curve matches reasonably the emission pattern at the two ends.
When viewed at an inclination of 60 degree, this curve, as noted above, would correspond to the Keplerian motion around a central mass of $\sim$ 4.8\solarmass.

\subsection{Radiative transfer modeling}

To aid the interpretation of observed P-V signatures, we pursue a set of radiative transfer model simulations using the "Simulation Package for Astrophysical Radiative Trans(X)fer" (SPARX) \footnote{https://sparx.tiara.sinica.edu.tw/}.
The software package produces image cubes through radiative transfer calculations of molecular line radiation with input physical models.
{We further convolve these synthetic image cubes output by SPARX with the observing resolution (0\farcs14) using the {\it "imsmooth"} command in CASA and generate the model P-V diagrams with the CASA {\it "impv"} task.}
We note that no interferometric spatial filtering was applied to form the model cubes during this process as we do not expect significant missing flux within the angular scale ($\sim$ 1\arcsec) of the observed structure.

For the physical model setup, we first assume that the emission originates from a flattened two-dimensional structure, possibly a disc or an envelope.
This is consistent with the elongation we see in the continuum as well as in the \mtcn\ integrated emission.
For the molecular gas column density and temperature, we adopt the empirical profiles inferred in Section 3.3. The column density radial profile is scaled to match the the observed gas column density or equivalently the observed (dust) continuum intensity with the gas temperature and a gas-to-dust ratio of 100 factored in.
We then set a \mtcn\ fractional abundance be $2\times 10^{-8}$ as suggested by Figure \ref{fig:fitplot} and discussed earlier.
The full size of the system is not fully known but we truncate the \mtcn\ abundance at a radius of 900~au ($\sim 0\farcs5$) based on the extent of the \mtcn\ emission observed in the P-V diagram. 
We carry out our calculations in the LTE regime since the observed transitions are most likely thermalized.

\begin{figure*}
\epsscale{1.2}
\plotone{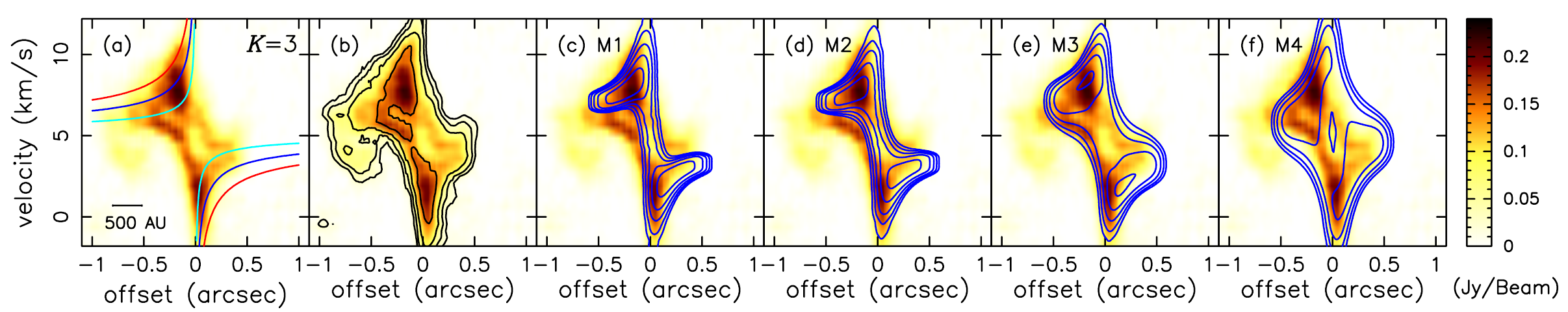}
\caption{\label{fig:figpvcmp} The position-velocity diagram of S255IR SMA1 in the \mtcn\ $J = 19-18$ $K$=3 transition along the P.A. of 165 deg in false color in all panels.
The red, blue, and green curves in panel (a) depict respectively the velocity profiles of Keplerian rotation of (8.04/$sin^2 i$) \solarmass, (3.57/$sin^2 i$) \solarmass, and (0.89/$sin^2 i$) \solarmass, where $i$ is the inclination angle of the system.
The black contours in panel (b) represent the observed emission at levels of  3, 6, 12, and 24 $\times$ 6mJy beam$^{-1}$ {($= 2 \sigma$)}.
The blue contours in panel (c), (d), (e), and (f) represent the modeled emission for Model 1, 2, 3, and 4, respectively, also at the same levels as in panel (b). {See Section 3.5 for more details.}}
\end{figure*}

In Figure \ref{fig:figpvcmp} we compare the observed $J=19-18, K=3$ emission features (shown in false color in all panels and in black contours in panel (b)) with the modeling results. 
In toy Model~1 (M1) and Model~2 (M2), we assumed the emissions are from gas following pure Keplerian rotation.
The blue contours in panel (c) for M1 and panel (d) for M2 display the P-V patterns with their central YSO masses and inclination angles being 20 \solarmass and 4.76 \solarmass, and 25 degree and 60 degree, respectively.
As one can see, the outer most contours of the model emission traces reasonably well the extent of the observed emission at the same level, suggesting that the empirical density, temperature, and abundance values helped gauging the simulation intensity be in the right ballpark.
There are slight differences between the two model patterns at small radii, but they generally agree with each other and resemble the observed features.
Nevertheless, there are subtle sub-structures that are not well reproduced.
In particular, the central emission "waist" between the two ends at the two "quadrants" in the simulated P-V pattern is significant narrower than that of the observed emission.
The observed wider "waist" features can be understood as signatures of infalling motions, which have been modeled and demonstrated extensively for low mass YSO systems by, for example, \cite{Ohashi97, Yen13, Sakai14a}.
{Such infalling features have also been recently outlined at high angular resolution toward, for example, the massive YSO G023.01--00.41 (see Fig. 5 of \citet{Sanna19}).}
In panel (e), we subsequently explore an infalling-rotation envelope model in which the gas moves in the azimuthal direction following the Keplerian manner (around a central mass of 4.76 \solarmass at an inclination angle of 60 degree) but also bears a radial infall velocity of 2.5 \kms at all radii.
The P-V emission pattern from this envelope model (M3), shown in Figure \ref{fig:figpvcmp} panel (e), captures the observed features well.
In particular, the infall motion leads to a wider "waist" in the P-V diagram.
Meanwhile, as opposed the M1 and M2 cases in which the emission peaks in the "central spine" of the "waist", the emission is stronger on the "edges" of the "waist", similar to that in the observed pattern.
In Figure \ref{fig:figpvcmp} panel (f), we furthermore hypothesize another model (M4) in which the gas follows the accretion flow prescribed by \cite{Ulrich76}.
The essence of such an accretion flow, which has been adopted for studying the envelopes of low- and high-mass stars forming regions in, for example, \citet{Yen14} and \citet{Keto10}, is that the gas parcel, while accelerates toward the central YSO due to the gravity, also have a rotational motion along a (ballistic) trajectory with its angular momentum conserved.
In this scenario, the gas stream lines are dictated by two of the three parameters, the (specific) angular momentum, the central mass, and the so called ``centrifugal barrier radius", at which point the infalling material can not move further in but presumably would get shocked, lose angular momentum, and settle into Keplerian rotation orbits around the central mass.
Admittedly, we have no knowledge for this model setup about the (specific) angular momentum or the centrifugal barrier radius, which is, however, likely to be small and unresolved.
For the purposing of testing, we assume arbitrarily its centrifugal barrier at a small radius of 90 au (or 0\farcs05).
As the gas density and thermal structures are readily parameterized with the empirical profiles, we mainly adopt the velocity fields of the accretion flow.
In this case, as in model M3, the feathers matches those with the observations quite well - the emission pattern is wider at the "waist" and is stronger on the edge of that.
The parameters of all the four radiative transfer models are shown in Table \ref{tab:models}.

\begin{figure*}
\plotone{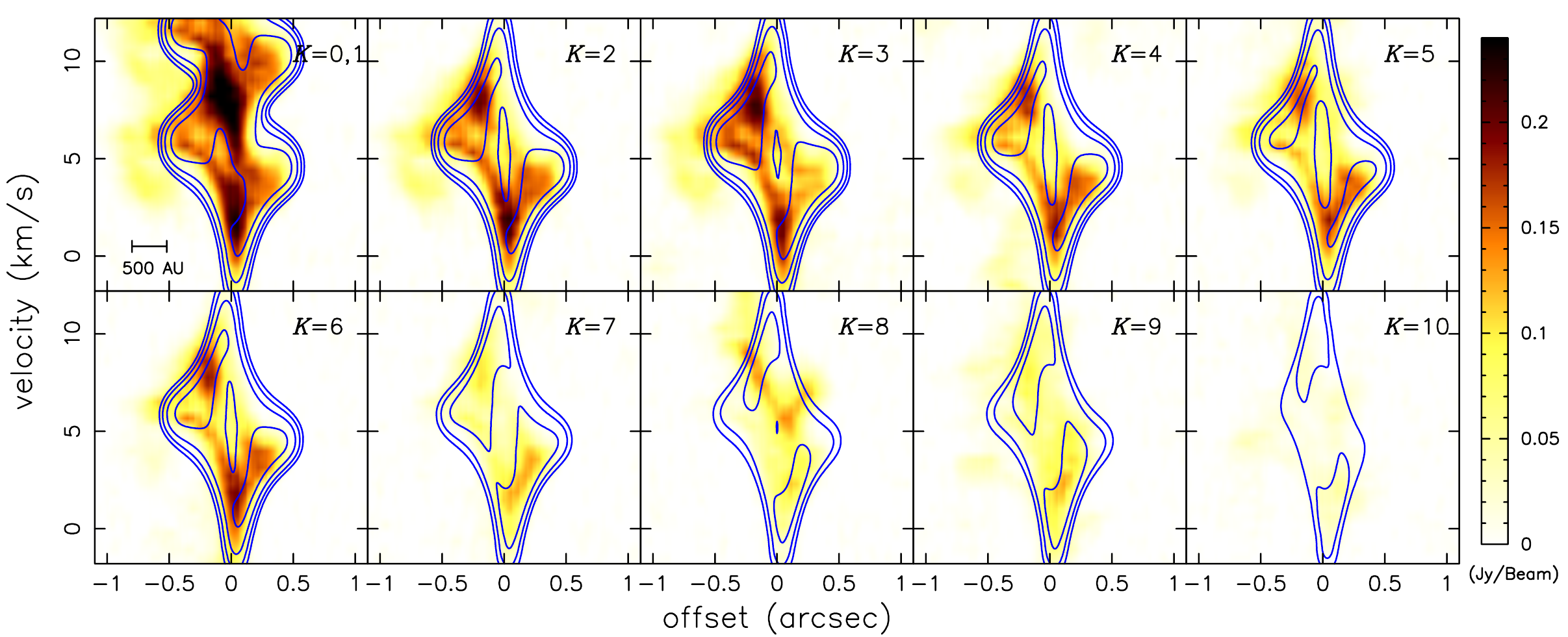}
\caption{\label{fig:figpvmodel} Similar to Figure \ref{fig:figpvplot}, the position-velocity diagram of S255IR SMA1 in the \mtcn\ $J$ = 19$-$18 $K$-ladder along the P.A. of 165 deg.
The observed intensities are shown in false color and the modelled intensities from Model~4 (M4) are shown in contours.
The contour levels are at 3, 6, 12, and 24 $\times$ 6mJy beam$^{-1}$ {($= 2 \sigma$)}, the same as Figure \ref{fig:figpvplot}. The $K$ component is labeled in the upper right corner of each panel and the length of a physical scale of 500 au is labeled in the top-left panel.}
\end{figure*}

In brief, models M3 and M4, with infalling gas motion factored in, exhibit better correspondence to the features seen in the P-V diagram as compared to models M1 and M2, in which only Keplerian motion is considered.
Given the reasonable match of M4, we present in Figure \ref{fig:figpvmodel} the P-V diagram of the observed \mtcn\ emission for all $K$ components in false color and the corresponding model emission in contours. 
These contours can be directly compared with those in Figure \ref{fig:figpvplot} and their consistencies can be recognized.

\section{Discussion} \label{sec:discussion}

\subsection{Central YSO mass and the geometry of the system}

To interpret the observed \mtcn\ kinematic signature, it is imperative to have some knowledge about the mass of {the central YSO NIRS3} and the geometry of the system.
Based on a distance of 2.5~kpc and the SED analysis, \citet{Ojha11} previously found 
%a luminosity of , which then translated to 
a mass of $\sim$ 27 \solarmass\ for the central exciting star in SMA1. 
By adopting an updated distance (of $\sim$ 1.6~kpc) hence a lower luminosity, \citet{Zinchenko15} revised the mass of the central YSO in SMA1 to be $\sim$ 20~\solarmass.
Indeed, considering the theoretical calculations by, for example, \citet{Hosokawa16} in which both stellar luminosity and accretion luminosity (with a mass accretion rate of $10^{-3}$ to 10$^{-4}$ \solarmass) are included, one would translate the observed luminosity of {S255IR~NIRS3/SMA1} at a level of at least $2.4 \times 10^{4}$ L$_{\odot}$ \citep[e.g.][]{Caratti17} to a mass of $\sim$ 15 -- 20 \solarmass\ for the central YSO.

While in Section \ref{sec:results} we have adopted an inclination angle of 60 degree for the radiative transfer models M2, M3 and M4, this angle in fact has not been very well constrained observationally.
According to the velocity gradient seen in the CH$_3$OH emission, 
\citet{Zinchenko15} previously inferred an inclination angle of $\sim$ 25 degree for an Keplerian disc surrounding the central 20~\solarmass YSO in SMA1.
The elongated morphology of the \mtcn\ emission seen by our observation, however, does not fully support such a scenario but instead would imply an inclination angle no less than 60 degree.
Incidentally, the mid-infrared (N band) observations with the MIDI instrument on the Very Large Telescope Interferometer (VLTI) carried out by \citet{Boley13} implied a highly inclined (79.6 degree) disc-like structure with an extension of 92.8 mas (or equivalently 165~au) in FWHM oriented at a P.A. of 149 degree based on their visibility fitting calculation with a two-dimensional Gaussian model.
The finite angular sampling and coverage in the $u$-$v$ plane (only between 32 degree and 76 degree) of their observations, however, might have limited the reliability of the measurement.
Besides, \citet{Burns16}, based on the maser kinematics measured both along the line-of-sight and in the sky plane through VLBI observations, suggested a jet from SMA1 inclined at 86 degree with a P.A. of 49 degree.
This jet geometry would imply also a highly inclined (86 degree) disc with a P.A. of around 140 degree, similar to that inferred from the VLTI MIDI observations.
In short, the orientation of the major axes obtained from the majority observations, including ours, are generally consistent.
Meanwhile, although the circumstellar disc has not been directly imaged, the inferred angle of such a structure is likely highly inclined.
The value of 60 degree assumed by us appears to be a lower limit.
{Since the YSO mass is reasonably between 15 -- 20  \solarmass, and the rotation axis is
inclined by more than 60 degrees with respect to the line-of-sight (as also suggested by
the well collimated outflow driven by the central star), the assumption of Keplerian
rotation does not reproduce the observations properly and has to be revised.}

\subsection{Disc or envelope?}

\subsubsection{An observed non-Keplerian velocity profile}

We have observed the molecular gas kinematics immediately surrounding {the massive YSO S255IR~NIRS3 in the star forming core SMA1} and the P-V diagram suggests its spin-up motion toward the center.
Results from radiative transfer exercises of Model~1 and Model~2, with gas emitting from a pure Keplerian disc appear to mimic the observed velocity gradient pattern reasonably to the zeroth order.
The derived radial surface density profile with $r^{-1.1}$ dependence is very close to the radial dependence of $r^{-1.0}$ in the theoretical model for a standard geometrically-thin steady-state accretion disc {\citep{Keto10}}.
It is therefore appealing to regard our observed structure as a rotationally-supported accretion disc.
Indeed, the scale of the structure of a few thousand au is quite comparable to the discs previously reported in other systems \citep{Johnston15, Chen16, Sanna19}.
However, in the standard accretion disc model, the infalling gas would have a (negative) radial velocity significantly smaller than the Keplerian velocity. 
This is not the case in S255IR~SMA1.
In the radiative transfer Model~4, we have assumed that gas falls inward in ballistic trajectories. 
In Model~3, a non-negligible infall velocity of 2.5 \kms\ is incorporated.
Furthermore, the assumed central stellar masses in Model~3 and Model~4 are much smaller 20 \solarmass, implying the azimuthal velocity of the surrounding gas is highly sub-Keplerian.
Such evidence instead suggests that the rotating structure is not strictly centrifugally supported like those pure Keplerian discs seen around low-mass YSOs. 
Rather, we are witnessing a rotating infalling "envelope" close to the central YSO. 
Meanwhile, there appears no direct evidence for the presence inside this "envelope" of a rotationally supported disc, which is directly involved in the mass accretion onto the star and jet/outflow launching.
If it exists, it would be presumably unresolved with a radius less than 125~au.

Based on the model parameters adopted in the radiative transfer models, we can further infer the mass {infall rate} of this envelope. 
In Model~3, the fact that we have the surface 
density profile with a radial dependence $r^{-1.1}$, which is very close to $r^{-1}$, and a fixed infalling velocity would gives rise to a nearly constant mass {infall} rate.
One can infer from, in this case, a surface density of 0.71 g cm$^{-2}$ at 900~au and an infall velocity of 2.5 \kms\ the mass {infall} rate to be 2.3 $\times$ 10$^{-4}$ \solarmass\ yr$^{-1}$.
Model~4 gives a modeled gas infalling velocity 2.99 \kms\ at 900~au and hence a comparable mass {infall} rate.
While there remain uncertainties in the mass {infall/accretion} rate measurements, {the} inferred mass {infall} rate from our observations of a few $\times$ 10$^{-4}$ \solarmass\ yr$^{-1}$ is significantly smaller than the {accretion} rate of 5 $\times 10^{-3}$ \solarmass yr$^{-1}$ estimated from the luminosity at burst by \cite{Caratti17}.

\subsubsection{Angular momentum considerations}

We judge that in Figure \ref{fig:figpvcmp} panel (a) the blue curve  represents the rotational velocity of the envelope reasonably relatively better than the other two (green and red) curves. 
This blue Keplerian velocity pattern, at an inclination angle of 60 degree, corresponds to a central mass no larger than $\sim$4.76 \solarmass.
Given that the central massive YSO is considered to have 20 \solarmass, the gas is is moving azimuthally in a sub-Keplerian manner.

In Model~4, the infalling rotating gas motion, due to the assumption of angular momentum conservation, should bear an toroidal velocity profile being inversely proportional to the radius, which is steeper than a Keplerian profile.
The absolute value of the azimuthal velocity as a function of radius depends on the the initial angular momentum and the central (stellar) mass, or the specific angular momentum.
If viewed edge-on, for example, the gas in the mid-plane would have an toroidal/azimuthal velocity $v_{\mbox{az}}$ of
\begin{equation}
 v_{\mbox{az}} = \frac{L}{m}\frac{1}{r} = L_{\mbox{\tiny S}}\frac{1}{r}
\end{equation}
where $L$ is the initial angular momentum, m is the mass of the infalling gas parcel, and $L_{\mbox{\tiny S}}$ is the specific angular momentum. 
The infalling gas in principle will reach the so called centrifugal barrier $r_{\mbox{\tiny CB}}$, where the gas kinetic energy are all in the toroidal motion.
The gas is assumed to get shocked, losing angular momentum and energy, and settle into Keplerian motion at this radius.
The centrifugal barrier $r_{\mbox{\tiny CB}}$ can be expressed as 
\begin{equation}
    r_{\mbox{\tiny CB}} = \frac{1}{2GM}(\frac{L}{m})^2 = \frac{1}{2GM}(L_{\mbox{\tiny S}})^2
\end{equation}
As noted just earlier, we do not detect any direct evidence of a transition between the a rotationally supported disc and a infalling envelope, the centrifugal barrier, if exists in our case, is likely within the central beam.
If we assume a central stellar mass $M$ of 20 \solarmass\ and an upper limit of $r_{\mbox{\tiny CB}}$ of 125~au (half of our spatial resolution 250~au), we reach an upper limit for the specific angular momentum $L_{\mbox{\tiny S}}$ of $\sim 3.6 \times 10^2$~au \kms\ in our observed infalling envelope.
Perpendicular to the outflow direction, \citet{Wang11} detected a velocity gradient ($\sim$ 2 \kms\ based on their Figure 14) in C$^{18}$O across a 10\arcsec-scale. 
Given the size scale of $\sim 2\times10^4$~au for this emission, they interpreted the structure possibly as a rotating toroid, which is fragmented into a multiple system (including SMA1 and SMA2).
Based on \cite{Zinchenko20}, the molecular gas traced by C$^{34}$S $J$=7--6 emission, on the other hand, bears a velocity gradient ($\sim$ 4 \kms\ between P2 and P4 their Figure 1) across a 4\arcsec-scale around SMA1.
The ``apparent'' specific angular momentum deduced from these velocity gradients are on the order of $\sim 10^4$ au \kms.
If our observed infalling gas at 1000~au scale originates from 10${^4}$~au scale traced by C$^{18}$O or C$^{34}$S, a large fraction of the angular momentum has to be dissipated during the process.

{In the context of angular momentum dissipation and the flatten, rotating and infalling "envelope" (or "sub-Keplerian disc") we uncover in S255IR~SMA1 as discussed in the last section, interstellar magnetic field may play crucial roles. Radiation magnetohydrodynamical simulations for massive star formation have demonstrated various effects due to the interplay between radiation, magnetic fields, rotation, and turbulences \citep[e.g.][]{Commercon11, Seifried11}. In the weak magnetic field regime, for example, centrifugally supported discs form with extents over 100 au. On the other hand, relatively stronger magnetic field would lead to more effective magnetic breaking, consequently transporting angular momentum outwards and suppressing the formation of Keplerian discs in the early stage of star formation \citep{Seifried11}. Such effects have been invoked by \cite{Sanna19} for interpreting the "sub-Keplerian disc" they imaged around G023.01--00.41 and we could observe a similar phenomenon. Future polarimetric observations for constraining the magnetic field morphology and strength in S255IR~SMA1 from hundreds to 10 thousand au scale will shed more light on the role of magnetic field in this region.
}

\subsubsection{Comparison to the burst mode of accretion in massive star formation}

As introduced earlier, {bf S255IR~NIRS3/SMA1} recently experienced a luminosity burst most likely associated with a disc-mediated accretion event.
Observations in the low-mass regime of star formation revealed that the infalling material lands onto a centrifugally supported accretion disc instead of directly interacting with the stellar surface.
While continuous mass inflow from the envelope to the inner circumstellar region replenishes the disc, the disc itself may be subject to gravitational instability.
The accretion of infalling disc clumps onto the protostellar surface triggers a sudden increase of the accretion rate and provokes in turn accretion-driven bursts since the debris are tidally stretched and get destroyed as they become closer to the vicinity of the growing YSO.
Such disc-based accretion bursts have been developed for the evolution of pre-main-sequence low-mass stars \citep{Vorobyov09,Elbakyan19}.
%The recently reported variability of maser emission monitored from the vicinity of high-mass protostars strengthens the suspicion of episodic mechanism occurring also in massive star formation \citep{Brogan18,Burns20,Chen20b}.
{In this context, while (periodic) maser variability associated with high-mass protostars has long been known \citep[e.g.][]{Goedhart03, Goedhart04}, only recently a number of prominent maser flare events have been directly correlated with accretion bursts in the vicinity of massive YSOs, suggesting new similarities to the low-mass star formation scenario \citep{Caratti17, Hunter17, Moscadelli17, Brogan18,Burns20,Chen20b}
}

To further interpret our observational signatures in the context of accretion bursts, we resort to a high-resolution gravito-radiation-hydrodynamics numerical simulation of the surroundings of a massive protostar. 
We initialize the model with the gravitational collapse of a 100~\solarmass\ rigidly-rotating pre-stellar core.
It has a temperature $T_{\rm c}=10$~K and a density distribution $\rho(r)\propto r^{-3/2}$, with $r$ being the radial coordinate. 
Its kinetic-by-gravitational energy ratio taken to be $\beta=4~\%$. 
We perform the simulation using a spherical midplane-symmetric computational domain $[r_{\rm in},R_{\rm c}]\times[0,\pi/2]\times[0,2\pi]$ of inner radius $r_{\rm in}=20~{\rm au}$ that extends up to $R_{\rm c}=0.1~{\rm pc}$, respectively, and we set $N_{\rm r}= 512 \times N_{\rm \phi}= 81 \times N_{\rm \theta} = 512$ grid cells. 
The inner boundary ($r\le20~{\rm au}$) is a sink cell localized at the origin of the domain, hence, material lost through the inner hole permits to calculate the accretion rate onto the protostar. 
We run the simulation up to $t_{\rm end}= 40.0~{\rm kyr}$ after the beginning of the collapse.
The system is integrated by solving the equations of gravito-radiation-hydrodynamics with the {\sc pluto} code\footnote{http://plutocode.ph.unito.it/}~\citep{Mignone07,Mignone12}.
The gas inertia is included in the calculations as described in~\citet{Meyer19a}.
The effects of the massive protostar's gravity is estimated with the total gravitational potential, and, additionally, our model also include the self-gravity of the circumstellar gas~\citep{Meyer18}. 
The direct stellar irradiation feedback of the central star~\citep{Meyer19b} and the radiative transport close to the star~\citep{Vaidya11} are both treated within the so-called gray approximation~\citep{Kolb13}. 

\begin{figure*}
\plotone{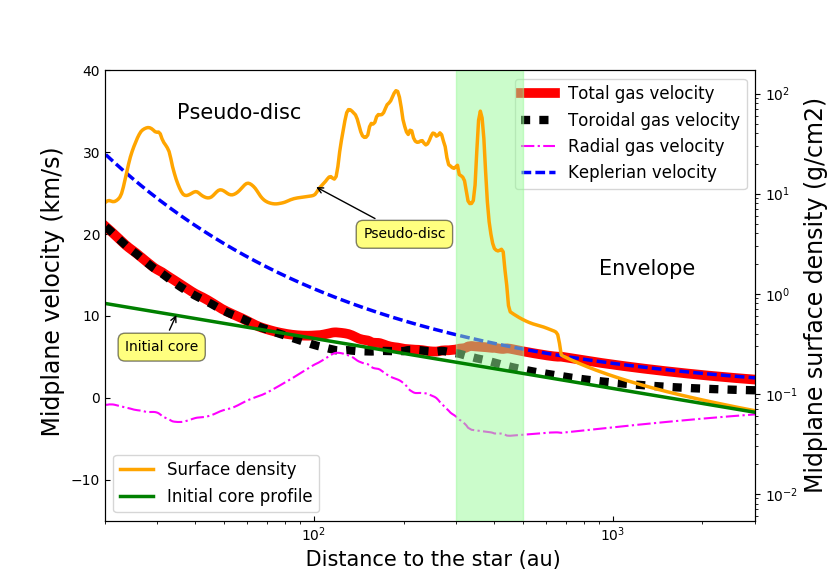}
\caption{\label{fig:figsim}
Azimuthally-averaged midplane gas velocity and surface density profiles of the close environment of a simulated massive protostar. 
The total gas velocity, the toroidal (gas) velocity, and the radial (gas) velocity are shown by the thick red solid line, the thick black dashed line, and the thin magenta dot-dashed line.
The initial pre-stellar core density profile is represented by the
solid green line and the surface density of the circumstellar medium is depicted by the solid orange line.
Yellow boxes are labels distinguishing density profiles from velocity profiles. 
The Keplerian profile calculated assuming a 20~\solarmass\ central protostar is represented by the dashed blue line. 
The vertical green-shaded region distinguishes the inner region with infalling sub-Keplerian gas,  
which density deviate from the pre-stellar core initial conditions, from the 
outer, still-infalling Keplerian envelope region, respectively. 
}
\end{figure*}

We select a simulation output on the basis of a total gas velocity profile consistent with the Keplerian velocity profile of a 20 \solarmass\ protostar.
That is the mass of {the central YSO in S255IR SMA1} inferred from the luminosity arguments.
At this time instance, 18.52~Myr after the onset of the gravitational collapse of its pre-stellar core, the free-fall gravitational collapse has ended.
Figure \ref{fig:figsim} displays the azimuthally-averaged midplane surface density and gas velocity profiles of the immediate surrounding environment of the simulated massive protostar.
The initial pre-stellar core density profile and the surface density of the circumstellar medium are shown by the solid green line and solid orange line, respectively.
In the same Figure, the total gas velocity, gas toroidal/azimuthal and radial velocities are shown, respectively, by the thick solid red line, thick dashed black line, and thin magenta dot-dashed line.
The Keplerian profile calculated assuming a $20\, \rm M_{\odot}$ central protostar is represented by the solid blue line.

There appears interesting and qualitative resemblances between the properties of the surroundings of 255IR SMA1 derived from the observations and those in the simulations displayed in Fig.~\ref{fig:figsim}. 
In the simulation, the density profile of the stellar surroundings is overall higher than the adopted initial pre-stellar core profile $\propto r^{-3/2}$ as the gas has been infalling.
A high-density disc-like structure, extending up to $\sim$ 600~au, has in particularly formed around the massive YSO.
This flattened structure, which is substantially enhanced in density but not rotationally supported, perhaps can be considered as an analog to the "inner envelope" or the "pseudo-disc" considered in low-mass star formation.
Its bumpy and rough density profile, meanwhile, implies the inhomogeneity in this component.
While the continuum and \mtcn\ emission in SMA1 extends to $\sim$ 900~au, we are likely seeing the same "inner envelope" or "psuedo-disc".
The azimuthally-averaged surface density profile at the edge of the inner envelope/psuedo-disc in the simulation is $\approx$ $0.5-1.0$~g cm$^{-2}$.
This appears to be at a comparable level to the surface density (0.71 g cm$^{-2}$) we infer at 900~au.
From the boundary of the pseudo-disc inwards to $\sim$ 200~au, the simulated gas motion bears a toroidal velocity at a sub-Keplerian level and a slow-varying (negative) radial velocity (of a few \kms).
Such rotating and infalling motion is roughly consistent with the velocity pattern we infer from the P-V diagram.
Limited by the angular resolution of $\sim$ 250~au, we are unable to resolve substructures within the very central region.
In short, S255IR~SMA1 appears therefore consistent with the burst mode picture of accretion disc scenario, in the sense that it might be surrounded by a pseudo-disc, nonetheless of slightly larger radius than that in the simulations. 
This discrepancy between observations and simulations might be explained by, amongst other caveats of the numerical modelling, the absence of magnetic field in the simulation.

We have estimated from our observations a mass {infall} rate smaller than {the accretion rate} inferred from the luminosity burst event.
The difference in these rates can be understood in the same context of episodic accretion.
What we estimated is the {infall} rate of the gas moving through the envelope,
while the luminosity burst reflects the variations of the accretion rate from very close stellar surroundings onto the stellar surface. 
This is largely consistent with the simulations of~\citet{Meyer17}. 
In their Fig.~2 one can see that the {mass infall} rate from the infalling envelope is a few $\times 10^{-4}$ \solarmass~yr$^{-1}$ (their red curve), while the corresponding peak in the accretion rate from the accretion disc onto the protostellar surface is about a few $\times 10^{-3}$ \solarmass~yr$^{-1}$ (their blue curve), respectively. 
This applies to the mild bursts the massive protostar undergoes, not to the strong FU-Orionis-like outbursts happening at times 17-23~Myr which are much powerful than the transient flare experienced by {bf the massive protostar S255IR NIRS3.}
{
%The rate inferred from the observations presented in this study is therefore in accordance with both (i) the burst mode of accretion in massive star formation developed in~\citet{Meyer17} and (ii) the constrain of 255IR's burst to be a rather small-amplitude burst~\citet{Meyer19a}, which has been similarly suggested by \citet{Liu18}.
%
Such mild bursts are likely common for massive YSOs to experience and they do not provoke dramatic changes in the stellar structure and radius~\citep{Meyer19b}. 
}

\subsection{Asymmetry and temporal behavior}

Numerical radiation hydrodynamical simulations of massive star formation such as the one introduced in the above section or those carried out in, for example, \citet{Klassen16} and \citet{Meyer17} suggest that the formation of disc, as it evolves, may appear highly asymmetric with clumps and/or spiral forms due to gravitational instability at hundreds to even thousands of au scales.
Gas clumps at these scales subsequently lose angular momentum and fall in towards 10s au scales and finally accrete on to the central massive YSO, giving rise to markedly elevated luminosity. 
While we do not witness direct signatures of spiral arms, our \mtcn\ data do suggest inhomogeneity and asymmetric kinematic features in the envelope, such as the asymmetric emission pattern seen in the P-V diagram.
The limited physical resolution of $\sim$ 250~au prevented us from revolving and identifying the small (10s~au) scale gas clumps that might have been physically associated with or left over from the most recent burst event.

\cite{Johnstone13} computed the temperature and luminosity variation of deeply embedded (low mass) stars due to episodic accretion events. 
They demonstrated that the dust temperature responses much quicker than the gas temperature to luminosity variations.
In particular, at the outer envelope where the dust emission is optically thin, heating of the gas would take year-long timescale as compared to that (month-long timescale) for the dust.
For S255IR~SMA1, evidences show that the burst occurred in 2015 has been waning down \citep{Liu18, Szymczak18, Uchiyama19}.
While our \mtcn\ data from the 2017 Sep (post burst) epoch do not show changes in their kinematic signatures, there appear tentative hints that the line intensities may have tailed off, suggesting reduced gas temperature in the region.
High resolution follow-up observations of \mtcn\ in the post-burst phase shall better examine this possible signature of gas temperature variation.

\subsection{Molecular gas mass estimates}

We have estimated the molecular column density and gas mass toward the S255IR~SMA1 by using the dust continuum emission. 
There exist a few potential caveats in inferring the gas mass due to the uncertainties in the assumed dust temperature, opacity, and measured flux.
As shown in \citet{Liu18}, the continuum flux of this region was varying during the luminosity burst.
The changing continuum flux is presumably reflecting the changing temperature, or even possibly dust properties including its opacity.
Only when the flux, temperature, and dust opacity are secured will we able to confidently measure the column density.
In the case of the massive star forming region G351.77-0.54, the gas temperature derived from the \mtcn\ $J = 37 - 36$ $K$-ladder was different from the dust temperature inferred from the dust continuum \citep{Beuther17}.
It was suggested by the authors that the dust continuum emission may trace the deeper and colder region while the \mtcn~emission originates from the hotter surface layer.
In our case, it is also possible that the gas and dust temperatures get de-coupled during a burst event as suggested by \citet{Johnstone13}.
{In addition, we have adopted the dust absorption coefficient $\kappa_{\nu}$ of 0.010 cm$^{2}$g$^{-1}$ based on \cite{Hildebrand83} and $\beta = 1.8$ from \cite{Chen16}. The index $\beta$ is, however, highly uncertain --- a change of $\beta$ to a smaller value of 1.5 or a larger value of 2 would, respectively, lead to an increase of $\kappa_{\nu}$ by 46\% or a decrease of $\kappa_{\nu}$ by 23\%.
Alternatively, if we consider  the value of dust absorption coefficient for "naked" dust at a density of 10$^8$ cm$^{-3}$ as tabulated  in Table 1 of \cite{Ossenkopf94} and again a gas-to-dust ratio of 100, $\kappa_{\nu}$ would increase by over a factor of 10 to be  0.113 cm$^{2}$g$^{-1}$, thus reducing the mass of the infalling envelope and increasing the \mtcn\ abundance within by the same factor.}
Finally, the observed flux variation in the 850~$\mu m$ continuum in the luminosity burst is about a factor of 2 \citep{Liu18}.
Considering all these aspects, the uncertainly in the mass estimate could be a factor of a few.
The currently derived gas mass of {around 1.64 \solarmass}\ within $\sim$ 2700~au appears significantly less than the presumed central stellar mass of $\sim$ 20 \solarmass, and hence the subsequent model calculations of gas kinematics without considering their self-gravity should be reasonable.

\subsection{Chemical signatures}

Chemical differentiation in between the circumstellar disc and envelope components has been seen toward low-mass YSOs, and it is perhaps best exemplified by the case of the Class~0 YSO L1527.
\citet{Sakai14a, Sakai14b} found that toward L1527 the (cyclic-)C$_3$H$_2$, CCH and CS emissions are doubly peaked and tracing the infalling rotating envelope.
On the other hand, the SO, H$_2$CO, and CH$_3$OH emissions are centrally peaked, tracing primarily the rotationally supported Keplerian disc component inside the centrifugal barrier.
\citet{Lee19}, by vertically resolving the almost edge-on protostellar disc in the low-mass YSO HH212, reported that complex organic molecules such as CH$_3$OH, CH$_3$CHO, CH$_3$CH$_2$OH, as well as H$_2$CCO, HCOOH, and CH$_3$OCHO, normally considered as tracers of hot molecular cores and corinos, seem to reside in the warm disc atmosphere within the centrifugal barrier.

Similar signatures of chemical differentiation have also been indicated toward massive YSOs.
For example, in the study of G339.88-1.26 by \citet{Zhang19}, there appear distinct spatial extent and kinematic signatures in the position-velocity diagrams for different species.
In this study, CH$_3$OH and H$_2$CO are tracing a more extended infalling rotating envelope while SO$_2$ and H$_2$S are most likely tracing a disc inside and particularly enhanced at the centrifugal barrier \citep{Zhang19}.

For the case of S255IR SMA1, we have detected and listed a good number of species/transitions in Table~\ref{tab:trans}, {as noted in Section 3.2}.
{The majority features seem} to all trace the same full extent of the envelope we see.
Partially blended CCS and SO$_2$ emission features are detected at around 349.225 GHz within the band.
While we do not find firm evidences of differentiation in terms of their patterns in the P-V diagram (see Fig.~\ref{fig:figfullpv}), there are hints that the CCS and SO$_2$ emissions may originate more prominently from smaller radii, which appears somewhat similar to the case of G339.88-1.26.
Details of the full chemical signatures in this region will be further discussed in a forthcoming study, and observations with higher sensitivity are required to verify the case of chemical differentiation.

\section{Summary}

We have carried out high angular resolution observations with ALMA toward the {massive star forming core} S255IR~SMA1 and imaged its 900 $\mu m$ dust continuum and the \mtcn\ $J$=19$-$18, $K$=0 to $K$=10 transitions amid rich spectral features.

\begin{enumerate}
    \item We estimate from the continuum emission {a molecular hydrogen density of 6.6 $\times$ 10$^{9}$ cm$^{-3}$} toward its most inner 250 au region, and {a total gas mass of 1.64 \solarmass\ within 1\farcs5 ($\sim$ 2700 au)} immediately {surrounding S255IR~NIRS3 in the core}.
    \item We find that the \mtcn\ emission is optically thick in the lower K transitions. The integrated \mtcn\ emission exhibits an elongated feature in the northwest-southeast direction with an extent of 1800 au and a position angle of 165 degree.
    \item From the \mtcn\ profile fitting, we reveal the spatial distribution of the gas column density, temperature, kinematics, and \mtcn\ abundance. The gas surface density shows a radial dependence of $r^{-1.1}$. 
    The gas temperature reaches more than 400K toward the \mtcn\ emission peak, which coincides with the dust continuum peak. It plateaus within 200 au and falls off with a radial dependence of $r^{-0.7}$. The \mtcn\ emission features a velocity gradient also along the northwest-southeast, which is nearly perpendicular to the bipolar outflow, and has its largest velocity dispersion {in the vicinity of S255IR~NIRS3/SMA1}. The \mtcn\ fractional abundance is estimated to be 2 $\times$ 10$^{-8}$.
    \item Based on the modelling with radiative transfer calculations the kinematics of the \mtcn\ gas seen in the well resolved position-velocity diagrams, we infer that the \mtcn\ emission is best described by a flattened rotating envelope with infalling motion {(also referred to as a sub-Keplerian disc of pseudo-disc)}. A mass {infall} rate of a few $\times$ 10$^{-4}$ \solarmass\ per year is derived. The putative Keplerian disc directly involved in the mass accretion onto the star and jet/outflow launching, if exists, is likely smaller than 125~au.
    \item There appear qualitative resemblances between the gas properties (such as density and kinematics) of {the surroundings of the central massive YSO inferred from the current observations} and those in the numerical simulations particularly tailored for studying the burst mode of massive star formation.
    \item Tentative signs of temporal variation in the molecular emission and of chemical differentiation need to be confirmed with higher sensitivity observations.
 \end{enumerate}

\begin{deluxetable*}{rcCr}[b!]
\tablecaption{{List of Detected Molecular Transitions within the Observed Window} \label{tab:trans}}
\tablecolumns{4}
%\tablenum{2}
%\tablewidth{0pt}
\tablehead{
\colhead{Rest Frequency} & \colhead{Molecule} &
\colhead{Transition} & \colhead{E$_u$} \\ % & \colhead{Remark} \\
\colhead{(GHz)} & \colhead{} &
\colhead{Quantum Number} & \colhead{(K)} % & \colhead{}
}
\startdata
 349.47820470 & NH$_2$CHO & 16(2,14) - 15(2,13) & 153.0 \\ % &  (1) \\
 {\bf 349.45369990} & {\bf \mtcn} & {\bf J=19-18~K=0} & {\bf 167.7} \\ % & \\
 {\bf 349.44698670} & {\bf \mtcn} & {\bf J=19-18~K=1} & {\bf 174.9} \\ % & \\
 {\bf 349.42684970} & {\bf \mtcn} & {\bf J=19-18~K=2} & {\bf 196.3} \\ % & \\
 {\bf 349.39329710} & {\bf \mtcn} & {\bf J=19-18~K=3} & {\bf 232.0} \\ % & \\
 349.35457760 & g'Ga-(CH$_2$OH)$_2$ & 33(8,26)~\nu=1 - 32( 8,25)~\nu=0 & 309.2 \\ % & (2) \\
 {\bf 349.34634280} & {\bf \mtcn} & {\bf J=19-18~K=4} & {\bf 282.0} \\ % & \\
 349.32035150 & CH$_3$CHO & \nu_t=1~18(1, 17) - 17(1, 16), E & 369.3 \\ % & (3)  \\
 {\bf 349.28600570} & {\bf \mtcn} & {\bf J=19-18~K=5} & {\bf 346.2} \\ % & \\
 349.28077220 & \mtttcn & J=19-18~K=0 & 167.7 \\ % & (ynsu) \\
 349.27408660 & \mtttcn & J=19-18~K=1 & 174.8 \\ % & (4) \\
 349.25403260 & \mtttcn & J=19-18~K=2 & 196.2 \\ % & (-) \\
 349.23381500 & CCS    & N=27-26, J=26-25 & 239.6 \\ % & \\
 349.22705660 & SO$_2$ & 31(10,22)-32(9,23) & 700.7 \\ % & \\
 349.22061830 & \mtttcn & J=19-18~K=3 & 231.9 \\ % & (5) \\
 {\bf 349.21231060} & {\bf \mtcn} & {\bf J=19-18~K=6} & {\bf 424.7} \\ % & \\
 349.17385730 & \mtttcn & J=19-18~K=4 & 281.9 \\ % & (6) \\
 {\bf 349.12528710} & {\bf \mtcn} & {\bf J=19-18~K=7} & {\bf 517.4} \\ % & \\
 349.10695400 & \mtnl & 14(1,13)-14(0,14) -+ & 260.2 \\ % &  \\
 349.06569800 & CH$_3$OCHO & 28(9,19) - 27(9,18) A & 294.6 \\ % & (ynsu) \\
 349.06254320 & CH$_3$CHO & \nu_t=1~18(4,14) - 17(4,13), E & 398.3 \\ % & (7) \\
%- (8) tried NH2CHO, C2H5CN, sugar, HCOOCH3 but no good match
 349.05354190 & NH$_2$CHO & 19(3,17) - 19(2,18) & 220.7 \\ % &  (ynsu) \\
% \sout{349.0516} &  \sout{U} & & \\ % & (8) \\
 349.05165440 & NH$_2$CHO & 19( 3,17)- 19(2,18) & 220.7 \\ % &  (ynsu CDMS) \\
 349.04854000 & CH$_3$OCHO & 28(9,19) - 27(9,18) E & 294.6 \\ % & (ynsu) \\
 {\bf 349.02497050} & {\bf \mtcn} & {\bf J=19-18~K=8} & {\bf 624.3} \\ % & \\
 349.02081410 & NH$_2$CHO & \nu_{12}=1~16(2,14) - 15(2,13) & 568.5 \\ % & (9) \\
%- (9) NH2CHO v12=1, 349020.8141 MHz or c-C2H4O 348966.6180
 348.91500870 & CH$_3$OCHO & 28(9,20) - 27(9,19) A  & 294.6 \\ % & (ynsu) \\
 {\bf 348.91140120} & {\bf \mtcn} & {\bf J=19-18~K=9} & {\bf 745.4} \\ % & \\
 348.90948000 & CH$_3$OCHO & 28(9,20) - 27(9,19) E  & 294.6 \\ % & (ynsu) \\
 348.85557140 & CH$_3$CHO & \nu_t=1~18(4,15) - 17(4,14)  & 398.3  \\ % & (10) \\
 348.84704640 & g-CH$_3$CH$_2$OH  & 10(6,4)-9(5,4)~\nu_t=0-1 & 146.5 \\ % &  (11) \\
 348.84704640 & g-CH$_3$CH$_2$OH  & 10(6,5)-9(5,5)~\nu_t=0-1 & 146.5 \\ % &  (11) \\
 348.80258600 & g'Gg-(CH$_2$OH)$_2$ & 37(1,36)~\nu=0 - 36(2,35)~\nu=0 & 326.1 \\ % & (12) \\
 348.80067700 & g'Gg-(CH$_2$OH)$_2$ & 37(2,36)~\nu=1 - 36(1,35)~\nu=1 & 326.1 \\ % & (12) \\
 {\bf 348.78462450} & {\bf \mtcn} & {\bf J=19-18~K=10} & {\bf 880.7} \\ % & \\
 348.74927490 & CH$_3$CHO & \nu_t=2~18(3,16) - 17(3,15) A & 565.6 \\ % & (ynsu)  \\
 348.72056960 & g-CH$_3$CH$_2$OH & 4(4,1)-3(3,1)~\nu_t=1-0 & 89.6 \\ % & (13) \\
 348.71974920 & g-CH$_3$CH$_2$OH & 4(4,0)-3(3,0)~\nu_t=1-0 & 89.6 \\ % & (13) \\
 348.65229870 & g'G'g-CH$_3$CHOHCH$_2$OH (?) & 27(17,10) - 26(16,10) & 190.3 \\ % & \\
 348.65229870 & g'G'g-CH$_3$CHOHCH$_2$OH (?) & 27(17,10) - 26(16,11) & 190.3 \\ % & \\
 348.65229870 & g'G'g-CH$_3$CHOHCH$_2$OH (?) & 27(17,11) - 26(16,10) & 190.3 \\ % & \\
 348.65229870 & g'G'g-CH$_3$CHOHCH$_2$OH (?) & 27(17,11) - 26(16,11) & 190.3 \\ % & \\
 348.64469090 & \mtcn\ (?) & J=19-18~K=11 & 1030.0 \\ % & ? \\
 348.62925750 & S$_2$O (?) & 41( 2,39)-40( 3,38) &	409.3 \\ % & ? \\
\enddata
%\tablenotetext{a}{}
%\tablecomments{All detected/identified lines}
\end{deluxetable*}

%\clearpage

\begin{deluxetable*}{rcccc}[b!]
\tablecaption{Radiative Transfer Model Parameters  \label{tab:models}}
\tablecolumns{5}
%\tablenum{2}
%\tablewidth{0pt}
\tablehead{
\colhead{Model} & \colhead{Model~1} & \colhead{Model~2} & \colhead{Model~3} & \colhead{Model 4} \\
\colhead{Parameters} & \colhead{(M1)} & \colhead{(M2)} & \colhead{(M3)} & \colhead{(M4)}
}
\startdata
 Central Stellar Mass & 20 \solarmass\ & 4.8 \solarmass\ & 4.8 \solarmass\ & 4.8 \solarmass\ \\
 Inclination Angle & 25$^\circ$ & 60$^\circ$ & 60$^\circ$ & 60$^\circ$ \\
 Surface Density Profile & \multicolumn{4}{c}{$r^{-1.1}$} \\
 Surface Density (at 900au) & $2.0 \times 10^{23}$ cm$^{-2}$ & $1.8 \times 10^{23}$ cm$^{-2}$ & $1.8 \times 10^{23}$ cm$^{-2}$ & $1.8 \times 10^{23}$ cm$^{-2}$ \\
  & (0.79 g~cm$^{-2}$) & (0.71 g~cm$^{-2}$) & (0.71 g~cm$^{-2}$) & (0.71 g~cm$^{-2}$) \\
 Temperature Profile & \multicolumn{4}{c}{$ \left\{ \begin{array}{l@{\quad:\quad}l} 430~\mbox{K} & r < 214~\mbox{au} \\ 430~(r/214)^{-0.7}~\mbox{K} & r \geq 214~\mbox{au} \\ \end{array} \right\} $ } \\
 Velocity Profile & Keplerian & Keplerian & Keplerian + infall & Ulrich \\
 X[\mtcn] & \multicolumn{4}{c}{$2 \times 10^{-8}$} \\
 Turbulent Velocity & \multicolumn{4}{c}{2 \kms\ }
\enddata
%\tablenotetext{a}{}
%\tablecomments{All detected/identified lines}
\end{deluxetable*}

\acknowledgments

{We appreciate very much the anonymous referee for comments that greatly improve the clarity of the paper.
We thank Dr. Hsi-Wei Yen for fruitful discussions.}
S.-Y. Liu, Y.-N. Su, and K.-S. Wang acknowledge the support by the Minister of Science and Technology of Taiwan 
(MOST grants 108-2112-M-001-048 and 108-2112-M-001-052-).
IZ research was supported by the Russian Science Foundation (grant No. 17-12-01256).
This paper makes use of the following ALMA data: ADS/JAO.ALMA \#2015.1.00500.S. 
ALMA is a partnership of ESO (representing its member states), NSF (USA) and NINS (Japan), together with NRC (Canada), 
MoST and ASIAA (Taiwan), and KASI (Republic of Korea), in cooperation with the Republic of Chile. 
The Joint ALMA Observatory is operated by ESO, AUI/NRAO and NAOJ.
This research made use of the PLUTO code developed by A. Mignone at the University of Torino. 
The authors acknowledge the North-German Supercomputing Alliance (HLRN) for providing HPC resources.

%\vspace{5mm}
%\facilities{HST(STIS), Swift(XRT and UVOT), AAVSO, CTIO:1.3m, CTIO:1.5m,CXO}

%\software{astropy \citep{2013A&A...558A..33A},  
%          Cloudy \citep{2013RMxAA..49..137F}, 
%          SExtractor \citep{1996A&AS..117..393B}
%          }

\software{CASA \cite{McMullin07}, pluto \cite{Mignone07}, SPARX (https://sparx.tiara.sinica.edu.tw)}

\end{document}